\documentclass[pdflatex,sn-mathphys-num]{sn-jnl}

\usepackage{graphicx}%
\usepackage{multirow}%
\usepackage{amsmath,amssymb,amsfonts}%
\usepackage{amsthm}%
\usepackage{mathrsfs}%
\usepackage[title]{appendix}%
\usepackage{xcolor}%
\usepackage{textcomp}%
\usepackage{manyfoot}%
\usepackage{booktabs}%
\usepackage{algorithm}%
\usepackage{algorithmicx}%
\usepackage{algpseudocode}%
\usepackage{listings}%
\usepackage{tabularx} 
\usepackage{comment} 
\usepackage{longtable}
\usepackage{algorithm}
\usepackage{algpseudocode}
\usepackage{algpseudocode}
\usepackage{tikz}

\theoremstyle{thmstyleone}%
\theoremstyle{thmstyletwo}%

\theoremstyle{thmstylethree}%

\renewcommand{\algorithmicrequire}{\textbf{Input:}}
\renewcommand{\algorithmicensure}{\textbf{Output:}}

\begin{document}

\title[Article Title]{Uncertainty-Aware Estimation of Mis/Disinformation Prevalence on Social Media}


\author*[1,2]{\fnm{Ishari} \sur{Amarasinghe}}\email{iamarasinghe@uoc.edu}

\author[1]{\fnm{Salvatore} \sur{Romano}}\email{sromano1@uoc.edu}

\author[1]{\fnm{Jacopo} \sur{Amidei}}\email{jamidei@uoc.edu}

\author[3]{\fnm{Emmanuel M.} \sur{Vincent}}\email{emmanuel.vincent@feedback.org}

\author*[1,2]{\fnm{Andreas} \sur{Kaltenbrunner}}\email{andreas.kaltenbrunner@upf.edu}

\affil[1]{\orgdiv{Ethical Technologies and Connectivity for Humanity Research Centre}, \orgname{Universitat Oberta de Catalunya}, \orgaddress{\street{Rambla del Poblenou  154-156}, \city{Barcelona}, \postcode{08018}, \country{Spain}}}

\affil[2]{\orgdiv{Department of Engineering}, \orgname{Universitat Pompeu Fabra}, \orgaddress{\street{Tànger 122}, \city{Barcelona}, \postcode{08018}, \country{Spain}}}

\affil[3]{\orgdiv{Science Feedback}, \orgaddress{\street{21 Place de la République}, \city{Paris}, \postcode{75003}, \country{France}}}

\abstract{Estimation of mis/disinformation prevalence in social media is crucial for designing mitigation strategies to limit its impact. Yet, such estimations are subject to several uncertainties that are rarely quantified jointly. In this study, we present a methodological contribution in which confidence intervals were used to quantify uncertainties related to mis/disinformation prevalence. The analysis draws on a multi-platform, multilingual dataset annotated by professional fact-checkers. Data were collected between March and April 2025 from Facebook, Instagram, LinkedIn, TikTok, X/Twitter, and YouTube across four EU Member States  (France, Poland, Slovakia, and Spain). We account for different causes of uncertainty: (i) sample uncertainty, (ii) annotation uncertainty arising from human disagreement and misclassification, and (iii) data retrieval uncertainty induced by keyword-based data collection. First, we estimate the uncertainty arising from the different causes separately using confidence intervals, simulation-based methods, and bootstrapping.
Finally, we combined multinomial simulations of annotator behaviour with keyword and post-resampling to capture the joint impact of measurement uncertainty on mis/disinformation prevalence estimates. The proposed methodological approach highlights the importance of uncertainty-aware estimation of mis/disinformation prevalence for robust analysis. The empirical results of this study show that keyword-based data retrieval can exceed baseline variability, leading to wider confidence intervals around prevalence estimates.}


\maketitle

\section{Introduction}\label{sec1}

The increasing volume and rapid dissemination of mis/disinformation on social media constitute a significant societal challenge \cite{denniss2025social}. The prevalence of mis/disinformation can promote false beliefs, distort public discourse, and may facilitate the dissemination of propaganda \cite{ecker2024misinformation}. Mis/disinformation is a multifaceted and complex phenomenon \cite{EDMO_second_report} shaped by technological, social, and political dynamics \cite{Fallis}. Regulatory bodies have attempted to address this issue at a global scale through a combination of legislative and self-regulatory mechanisms. In the European Union (EU), the \textit{Code of Conduct on Disinformation}, first introduced in 2018 and substantially strengthened in 2022, provides a voluntary framework for fighting the spread of disinformation \cite{EC_CodePracticeDisinformation}. 

Major online platforms such as Google, Facebook, X (formerly Twitter), and Mozilla signed the Code in 2018, followed in subsequent years by additional organizations including Microsoft, TikTok, and the European Association of Communications Agencies (EACA) \cite{grabowskarole}. These commitments aim to enhance transparency, accountability, and evidence-based monitoring of platform practices. To evaluate the impact of the Code, the European Digital Media Observatory (EDMO) has proposed a set of Structural Indicators (SIs) designed to enable systematic and longitudinal assessment of disinformation across multiple dimensions \cite{EDMO_second_report}. Among these, prevalence (i.e., the proportion of the content that users are exposed to on the platform that contains mis/disinformation \cite{sfreport}) is identified as a foundational indicator, alongside indicators related to sources of disinformation, monetization of disinformation, and cross-platform aspects of disinformation \cite{EDMO_second_report}. It should be highlighted that among these SIs, prevalence, which \textit{aims to measure how widespread disinformation is across platforms}  \cite{EDMO_second_report}, is one of the most methodologically demanding indicators, as it requires inferring the frequency of mis/disinformation from large-scale, partially observed, and noisily labelled social media data.

Existing studies have attempted to quantify mis/disinformation prevalence using a range of approaches. Some rely on manual annotation of content, reporting descriptive prevalence estimates for specific platforms or topics (e.g., TikTok health-related content in \cite{glover2025fake}). Others estimate prevalence by manually annotating random samples and extrapolating proportions to a broader population, such as election-related Twitter polls \cite{scarano2025election}. In contrast, a substantial body of work focuses on mis/disinformation detection using machine-learning techniques, framing the problem primarily as a classification task rather than a measurement problem \cite{ding2025evolvedetector, liu2025systematic}. Recent studies have also used fact-checking corpora to study mis/disinformation prevalence across languages, using repeated or clustered fact-checks as indirect proxies for prevalence \cite{quelle2025lost}, while multilingual fact-checked datasets such as \textit{MuMiN} have been developed specifically to support classification rather than prevalence estimation \cite{nielsen2022mumin}. Across these studies, prevalence is most often reported as point estimates or descriptive proportions, with limited focus on uncertainty arising from data collection choices, partial annotation, and annotator disagreement, highlighting the need for uncertainty-aware prevalence estimations such as the work reported in this study.

In our work, we conceptualise mis/disinformation prevalence estimation as being subject to several causes of uncertainty. Specifically, we account for (i) sample uncertainty due to estimating prevalence from a subset of posts within a larger set of continuously evolving content on a social media platform, (ii) annotation uncertainty resulting from annotator disagreements, and  (iii) data retrieval uncertainty inherent in keyword-based data collection, where keyword choices lead to the inclusion of different subsets of posts.

We begin by estimating each uncertainty cause independently using confidence intervals, simulation-based approaches, and bootstrap resampling. We then combine keyword-level resampling with multinomial simulations of annotation outcomes to model the combined propagation of measurement uncertainty in prevalence estimates.

Our analysis is grounded in the European research project SIMODS (Structural Indicators to Measure Online Disinformation Scientifically)\footnote{https://recerca.uoc.edu/proyectos/888396/detalle}, which operationalises EDMO’s Structural Indicators (SIs) across six Very Large Online Platforms (VLOPs): Facebook, Instagram, LinkedIn, TikTok, X/Twitter, YouTube, and four countries: France, Poland, Slovakia, and Spain. 

The initial estimates of the SIs of SIMODS without uncertainty estimates have been published in a project report \cite{sfreport}. The present study extends this work by refining the statistical analysis and introducing, in particular, the uncertainty quantification for prevalence estimates in detail. By explicitly modelling and comparing different sources of uncertainty, this study aims to improve the robustness of prevalence indicators.

\section{Background and Related Work}\label{sec2}

This section first defines mis/disinformation and situates it in the EU’s Digital Services Act (DSA) framework. It then reviews empirical work highlighting why structural indicators are difficult to implement in practice, focusing on two dimensions—language and platform—and on the downstream implications for detection and prevalence estimation.

\subsection{Definitions and EU regulatory context}
The \textit{Code of Conduct on Disinformation} defines disinformation as “verifiably false or misleading information that is created, presented, and disseminated for economic gain or to intentionally deceive the public, and that may cause public harm” \cite{EC_CodePracticeDisinformation}. In this study, we do not differentiate between misinformation (i.e., false or misleading information shared unintentionally) and disinformation (where intention to deceive is present), as both might cause harm. Moreover, in practice, distinguishing intentional deception from unintentional sharing is rarely possible based on content alone and is difficult to operationalize reliably in automated or large-scale analyses \cite{sfreport}. Assessing intentionality typically requires contextual evidence about the producer’s goals, coordination, incentives, or prior behavior—signals that are often unavailable or noisy. Thus, in this study, we treat both forms under a unified definition while explicitly acknowledging that our assessment does not depend on proving intent. Throughout the study, we use the term \textit{mis/disinformation} to refer to both phenomena collectively.

On 13 February 2025, the European Commission and the European Board for Digital Services endorsed the incorporation of the 2022 Code of Practice on Disinformation into the DSA framework, thereby elevating it to the status of the Code of Conduct on Disinformation.
Although participation remains voluntary, commitments become enforceable once an actor signs up, as adherence to the Code can be used as evidence of compliance with DSA's risk-mitigation obligations and is subject to regulatory oversight by the European Commission. Non-compliance may therefore contribute to enforcement actions under the DSA’s supervision and sanctions regime~\cite{codeConduct}.

Substantively, the Code of Conduct is closely related to the DSA’s systemic-risk framework for Very Large Online Platforms (VLOPs) and Very Large Online Search Engines (VLOSEs). Articles 34 (on risk assessment) and 35 DSA (on risk mitigation) of the DSA require such providers to assess and mitigate systemic risks arising from the design, functioning, and use of their services, including risks to civic discourse, electoral processes, and public security—categories under which disinformation is explicitly understood to fall \cite{codeConduct}. 

In practice, assessing each platform’s disinformation-related actions through the transparency reports required under Article 15 DSA (transparency reporting obligations) is challenging due to repetition, vague descriptions, and limited data quality. For example, 40\% of reported actions have an unclear geographic scope and 68\% lack outcome metrics \cite{park2022beyond}.
In this context, independent audits and Article 40 DSA (on data access) are particularly important for scrutinizing and mitigating systemic risks linked to the spread of mis/disinformation. 
These mechanisms also play a fundamental role in ensuring platform accountability, as disregarding credible independent evidence of disinformation-related risks could constitute a failure to comply with the DSA’s risk-assessment obligations and may trigger scrutiny by the European Commission \cite{liesenfeld2025legal}. 

The Code of Conduct is intended to operationalise these abstract legal obligations by specifying concrete mitigation measures, cooperation mechanisms, and reporting practices related to disinformation. As such, adherence to the Code can serve as structured evidence that a platform has adopted appropriate and proportionate risk-mitigation measures within the meaning of Article 35 DSA. However, compliance with the Code does not remove the platform’s independent responsibility to fulfil its obligations under the DSA, particularly given the significant challenges that remain in the implementation phase \cite{brogi2024code}.

\subsection{Language as an implementation variable}

In our study, prevalence estimates and uncertainty analyses are reported at the level of language and platform-language combinations. First, considering language, it is not simply a descriptive characteristic of the dataset, but an important analytical dimension through which structural indicators are operationalised and interpreted. Treating language as an implementation variable is essential because our prevalence levels and uncertainty patterns may vary across linguistic contexts, as language influences keyword-based content retrieval, annotation practices, and ultimately prevalence estimates.

Research shows that around 33\% of repeated misinformation claims appear in more than one language \cite{quelle2023lost}, indicating that some narratives cross language borders. However, misinformation circulates much more frequently within the same language community than across languages \cite{quelle2023lost}. In other words, while some claims reach multiple linguistic contexts, most sharings and repetitions happen within a single language group. Moreover, cultural and linguistic differences also shape how emotions and rumours are expressed \cite{pranesh2021looking}. Together, these findings suggest that prevalence levels may differ across linguistic contexts and should not be assumed to be uniform. Existing studies have also shown that automatic classification of misinformation is also more challenging in multilingual settings, as models trained in one language often perform poorly in others \cite{khare2019relevancy,awal2022muscat}. Furthermore, outside the United States, reliable data on the scale of misinformation remains limited in many countries \cite{fletcher2018measuring}, which makes multi-lingual data collection essential. In this context, our study contributes by estimating the prevalence of mis/disinformation using a multi-lingual corpus. 

\subsection{Platform constraints, cross-platform comparability, and prevalence estimation}

Similarly, our results are presented at the platform and platform-language level, as platform-specific data access constraints directly influence what data can be collected and how prevalence is estimated. Since the 2018 Cambridge Analytica scandal, API (Application Programming Interface) restrictions have intensified, with platforms limiting or throttling data access \cite{awal2022muscat}. As a result, researchers often rely on alternative methods such as web scraping or browser extensions, which introduce their own biases \cite{awal2022muscat}. These differences in access and data collection practices complicate the meaningful standardisation and comparability of metrics across platforms \cite{rogers2021marginalizing}.

Existing studies have also shown that the same misinformation claims can appear on multiple platforms, but often in different formats, lengths, levels of detail, or even modalities. Moreover, Platforms have different affordances and norms, which shape how content is presented. As a result, detection approaches are often platform-specific and do not easily generalise across platforms \cite{panchendrarajan2024claim}. Research also shows that even when a claim appears on several platforms, its structure and presentation may differ substantially \cite{hale2024analyzing}. For our study, this means that content cannot be treated as fully comparable across platforms. Platform differences influence both retrieval and annotation processes, reinforcing the need to estimate prevalence separately at the platform and platform-language level.

Moreover, automatically detecting mis/disinformation across multiple social media platforms is difficult because the data available only captures a subset of content and does not fully represent the content circulating online. In this context, previous studies have shown that there is little overlap between users of different tiplines (people who choose to send suspicious content to a specific fact-checking organization), and that new users continue to introduce previously unseen content \citet{hale2024analyzing}. This suggests that the observed corpus is unlikely to be exhaustive, limiting the coverage of mis/disinformation on such datasets.

Finally, estimating the prevalence of mis/disinformation may constitute several statistical challenges. First, mis/disinformation typically constitutes a small fraction of total content online, which creates a low base-rate problem in which a small sampling fluctuation or classification error affects prevalence estimates. Second, prevalence estimates depend on other processes, such as keyword-based data retrieval and annotation by fact-checkers. Keyword-based retrieval influences which content is included or omitted from the sample, potentially introducing a selection bias. Human annotations are subject to interpretation differences and disagreement and can introduce bias. Our study, therefore, considers these challenges and model uncertainty caused by several factors, as described in the following sections.

\section{Methods} \label{sec3}

This study models multiple causes of uncertainty affecting mis/disinformation prevalence estimates. First sample uncertainty, which provides a baseline estimation, is quantified using confidence intervals. Specifically, we use the Wilson score interval to construct confidence intervals. Next, annotation uncertainty arising from annotator disagreements during the labeling process is modeled using a multinomial simulation approach. Then the data retrieval uncertainty introduced by keyword-based content collection is modelled through a bootstrap resampling procedure. Finally, we combine multinomial simulations and a bootstrap resampling procedure to capture the joint impact of these uncertainty causes. The following section first outlines data collection and pre-processing steps. Then, the modelling approach developed to systematically account for multiple causes of uncertainty in prevalence estimation is presented.

\subsection{Data Collection}

Data were collected from six VLOPs (Facebook, Instagram, LinkedIn, TikTok, X/Twitter, YouTube) and four EU Member States (France, Poland, Slovakia, Spain) between March  17 and April 13, 2025 using a platform-adaptive strategy that combined formal data access requests under the Digital Services Act (with only LinkedIn providing random samples) and keyword-based retrieval through native platform search functionalities or licensed third-party tools. Collection procedures were tailored to each platform’s technical constraints, with daily or biweekly searches as appropriate, post-hoc filtering by publication date where necessary, and subsequent updates of engagement metrics. Together, this approach yields a corpus that approximates the range and visibility of content to which users are exposed across platforms and countries \cite{sfreport}.

To approximate the information environment that users encounter online, a multilingual corpus was constructed using approximately 100 curated keywords per language covering five major topical domains, i.e., the Russo–Ukrainian conflict, climate change, health (including COVID-19), migration, and local politics. Keyword lists were developed and validated by professional fact-checkers to balance neutral keywords (representing about 50\%) with ambiguous and misinformation-related terms representing the other half of the keywords.

Because keyword-based collection can introduce topical noise (e.g., entertainment or celebrity-related content), a large language model (GPT-4o-mini) was used to support the manual filtering of irrelevant content (see the Prompt used in Appendix~\ref{app-Prompt}). The model assisted in identifying posts that were unlikely to relate to the target topic, enabling more efficient corpus refinement. This filtering step retained only posts contributing to public discourse, including topics such as health, science, politics, and climate, while excluding irrelevant material.

\subsection{Data Annotation and Pre-processing} \label{annotations}

Professional fact-checkers were involved in the manual data-annotation process. The annotated dataset consists of two parts: (1) a double-coded subset, in which each post was independently annotated by both a Junior (Jr) and a Senior (Sr) fact-checker, and (2) a Junior-only remainder, comprising posts annotated only by a Junior fact-checker.

Before the data annotation, pilot tests were conducted to refine the coding scheme, resulting in a final set of nine categories (summarised in Table~\ref{tab:labels}). To ensure the annotated data reflected the content with the greatest audience reach, a view-weighted random sample \footnote{https://pandas.pydata.org/docs/reference/api/pandas.DataFrame.sample.html} of 500 posts per platform for each country ($N=3,000$) was retrieved. This approach ensures that the probability of a post being selected is proportional to its number of views, thereby prioritising content with higher visibility in the corpus.

For each language, a Junior fact-checker annotated the full sample of 3,000 posts. In parallel, a Senior fact-checker independently reviewed a randomly selected 20\% subset of the same sample. Given the inclusion of six platforms, this resulted in 600 posts per language being annotated by both a Junior and a Senior fact-checker, forming the double-coded subset. The remaining posts constitute the Junior-only remainder.

After the initial annotation, a resolution phase (second round) was conducted in which disagreements within the double-coded subset were discussed, and a final agreed-upon label was assigned to each disputed item. Additional details on data collection and annotation procedures are provided in \cite{sfreport}.

\begin{table}[b]
\centering
\caption{Annotation categories and their definitions}
\label{tab:labels}
\begin{tabularx}{\linewidth}{@{}lX@{}}
\hline
\textbf {Label} &\textbf{Definition} \\
\hline
\textbf{Mis/disinformation} & Content stating or clearly implying a verifiably false or misleading claim that may cause public harm. \\

\textbf{Credible and informative} & Content conveying true or credible information on important matters about the state of the world (excluding trivia, gossip, or anecdotes). \\

\textbf{Borderline} & Content feeding a misleading narrative without necessarily containing outright falsehoods, but potentially reinforcing false beliefs. \\

\textbf{Abusive} & Content not containing mis/disinformation but involving harmful material such as hate speech, insults, spam, or incitement to harmful behaviour. \\

\textbf{Unverifiable} & Content that cannot be assessed as either credible or mis/disinformation (e.g., opinion-based). \\

\textbf{Irrelevant} & Content not about public affairs or scientific/political issues (e.g., entertainment, sports, religious content, cooking recipes without health claims, or geographically irrelevant to Europe). \\

\textbf{Other language} & Content not written in one of the target languages or in English. \\

\textbf{Deleted} & Content unavailable at the time of annotation (e.g., removed from the platform). \\
\textbf{Don’t know} & Content not fitting any other category. \\

\hline
\end{tabularx}
\vspace{2pt}
\end{table}

\begin{figure}[!ht]
    \centering
    \includegraphics[width=\columnwidth]{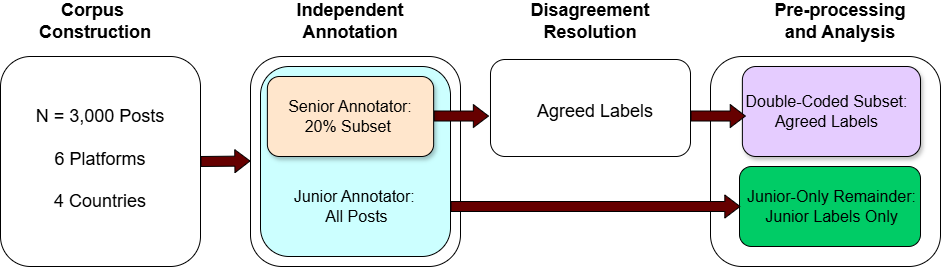}
    \caption{Overview of the Data collection and annotation process}
    \label{fig:annotation-processing}
\end{figure}

\begin{figure}[!ht]
    \centering
    \includegraphics[width=\columnwidth, height=0.3\textheight]{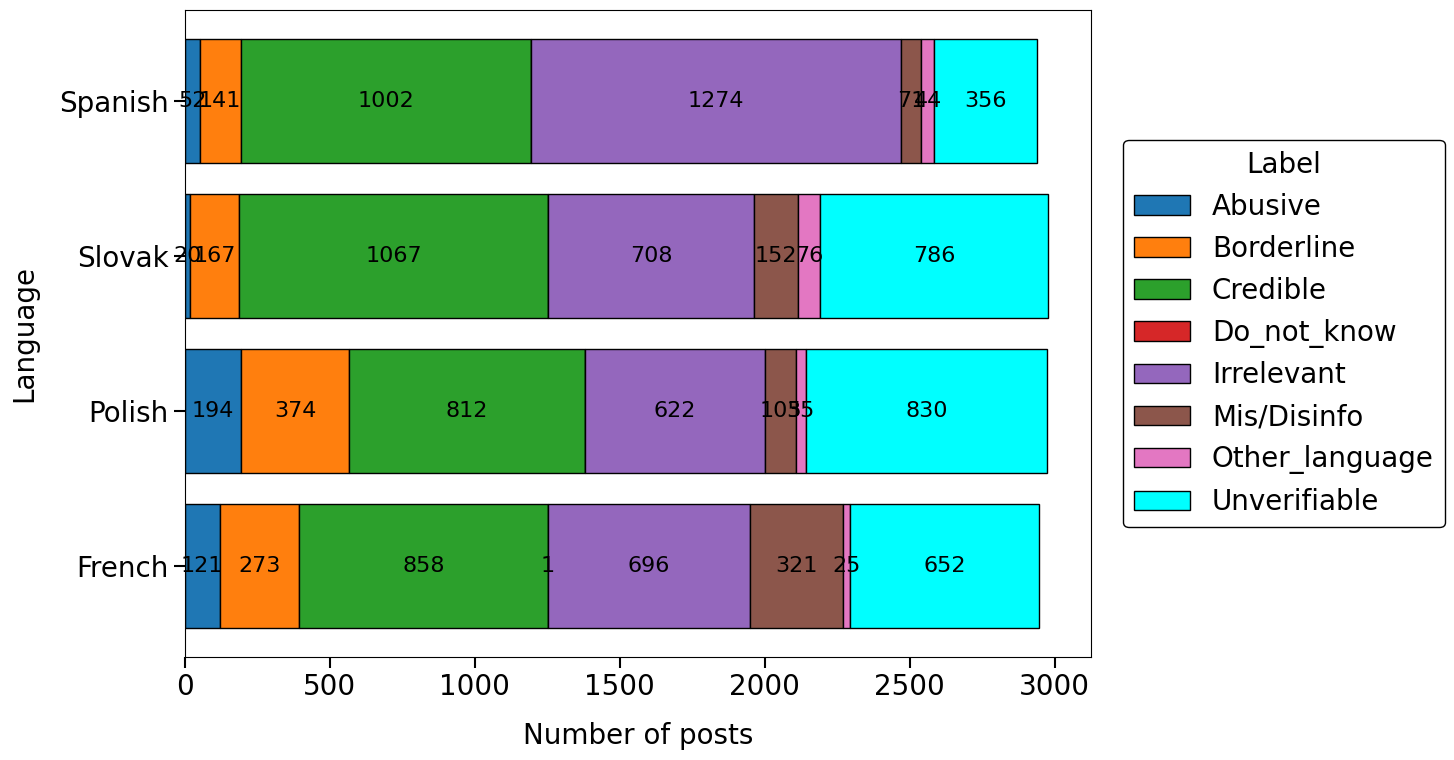}
    \caption{Distribution of Labels per Language in the pre-processed corpus}
    \label{fig:lang_distribution}
\end{figure}

\begin{figure}[!ht]
    \centering
    \includegraphics[width=\columnwidth, height=0.3\textheight]{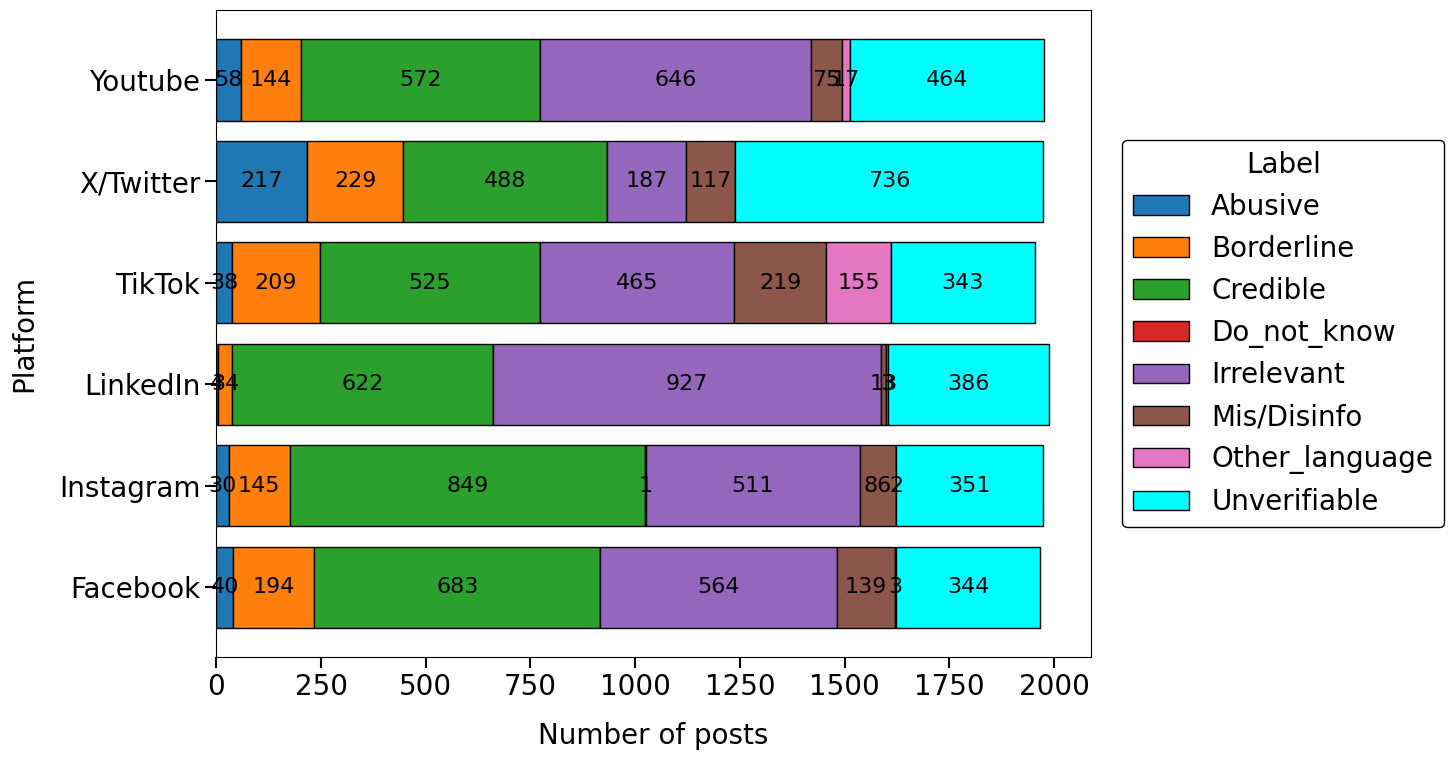}
    \caption{Distribution of Labels per Platform in the pre-processed corpus}
    \label{fig:plat_distribution}
\end{figure}

When processing the annotated datasets to estimate mis/disinformation prevalence, two data-quality issues were identified: First, some posts were labelled as \textit{Deleted} by both annotators, indicating that the content was unavailable at the time of annotation. Second, in some cases, posts were available to the Junior annotator but no longer accessible to the Senior annotator during the independent review, leading the latter to assign the \textit{Deleted} label. All such records were removed before analysis. Therefore, after preprocessing, the size of the double-coded subset was 592 posts for Polish, 581 for Spanish, 592 for French, and 595 for Slovak. The remaining posts were annotated only by Junior fact-checkers. This annotation process is summarized in Figure~\ref{fig:annotation-processing}. The final dataset sizes combining the double-coded subset and the Junior-only remainder after pre-processing were 2,947 posts for French, 2,972 posts for Polish, 2,976 posts for Slovak, and 2,940 posts for Spanish. The distributions of the labels per language and per platform of the pre-processed dataset are shown in Figure~\ref{fig:lang_distribution} and Figure~\ref{fig:plat_distribution}, respectively.

\subsection{Analysis Units and Aggregation Levels}

Prevalence estimates and associated uncertainty measures are computed at three levels of aggregation, referred to throughout the paper as \emph{analysis units}. For each analysis unit \( u \), prevalence is estimated using the corresponding subset of annotated posts. These units enable comparison across linguistic and platform contexts. 

Specifically, we consider:
(i) the \textbf{language level} (\( u = \ell \)), which captures prevalence within each language across platforms;
(ii) the \textbf{platform level} (\( u = p \)), which captures prevalence within each platform across languages; and
(iii) the \textbf{platform-language level} (\( u = (p,\ell) \)), which captures prevalence within specific platform--language combinations.

\subsection{Baseline Prevalence Estimates}

The baseline prevalence estimate provides an estimate of the proportion of mis/disinformation in the pre-processed corpus. Given that our corpus constitutes a random sample rather than the full population of posts, different samples drawn under the same procedure would result in slightly different estimates. This sampling variability is quantified using confidence intervals around the observed prevalence. In this study, we use the Wilson score interval \cite{Wilson01061927} to construct the confidence intervals for the binomial proportions \cite{brown2001interval}. As shown in Equation~\ref{eq:wilson}, let $x$ denote the number of mis/disinformation posts in a sample of size $n$, $\hat{p} = \frac{x}{n}$ is the observed prevalence and $z$ denotes the $(1-\alpha/2)$ quantile of the standard normal distribution (e.g., $z=1.96$ for a 95\% confidence interval). Compared to the standard Wald interval, Wilson score has been shown to provide better coverage properties when the sample size is moderate \cite{agresti1998approximate}.

\begin{equation} \label{eq:wilson}
\mathrm{CI}_{\text{Wilson}} =
\left(
\frac{
\hat{p} + \frac{z^2}{2n}
-
z \sqrt{
\frac{\hat{p}(1-\hat{p})}{n}
+
\frac{z^2}{4n^2}
}
}{
1 + \frac{z^2}{n}
},
\;
\frac{
\hat{p} + \frac{z^2}{2n}
+
z \sqrt{
\frac{\hat{p}(1-\hat{p})}{n}
+
\frac{z^2}{4n^2}
}
}{
1 + \frac{z^2}{n}
}
\right),
\end{equation}

Baseline estimates were computed using the pre-processed corpus, before accounting for additional sources of uncertainty, such as annotation disagreement or keyword-based data collection, which are addressed in the subsequent sections. Baseline prevalence, therefore, provides a reference point against which the impact of these additional uncertainty components can be assessed.

Two prevalence definitions can be used. For instance, mis/disinformation prevalence can be defined as the proportion of content labelled as mis/disinformation relative to content addressing similar topics that is assessed as legitimate (see Equation~\ref{eq:1}). In this study, we adopt this restricted prevalence as our primary definition of mis/disinformation prevalence, as it focuses on content for which a meaningful distinction between misleading and legitimate information can be made.

\begin{equation} \label{eq:1}
{P}_{\text{restricted}} = 
\frac{N_{\text{mis/disinfo}}}{N_{\text{mis/disinfo}} + N_{\text{cred}} + N_{\text{unverif}}} \times 100
\end{equation}

where \( N_{\text{mis/disinfo}} \), \( N_{\text{cred}} \), and \( N_{\text{unverif}} \) 
are the numbers of posts labeled as \textit{Mis/Disinformation}, \textit{Credible/Informative}, 
and \textit{Unverifiable}, respectively.\\

As a complementary measure, prevalence could also be defined as the proportion of posts labelled as mis/disinformation relative to all annotated posts (see Equation~\ref{eq:2}). Results based on this definition are reported in the Appendix (see Appendix~\ref{app-C}).

\begin{equation} \label{eq:2}
{P}_{\text{total}} = 
\frac{N_{\text{mis/disinfo}}}{N_{\text{total}}} \times 100
\end{equation}

\subsection{Uncertainty Estimates}

While baseline prevalence estimates quantify uncertainty due to finite sampling from the content corpus, they do not capture additional sources of uncertainty arising from the data collection and annotation process. In practice, prevalence estimates are also shaped by how content is retrieved and how labels are assigned, both of which introduce variability that is not reflected in standard confidence intervals.

In this study, we consider two additional sources of uncertainty that are particularly important for mis/disinformation monitoring, namely \textbf{annotation-related uncertainty} and \textbf{data retrieval uncertainty}. To capture these additional uncertainty components, we employ complementary approaches. Annotation-related uncertainty is modeled using a multinomial simulation that propagates uncertainty from the double-coded subset to posts annotated only by Junior fact-checkers. Data retrieval uncertainty is quantified using a two-level bootstrap procedure that reflects variability at both the keyword and post levels. Finally, we combine these components to derive joint uncertainty estimates, providing a more comprehensive characterization of uncertainty in mis/disinformation prevalence, as described in the following sections. 

\subsubsection{Annotation-related uncertainty} \label{annot_uncert}

\paragraph{Modelling Annotation Disagreement}

First, regarding the annotation-related uncertainty, it arises from disagreement between fact-checkers, reflecting the inherent difficulty of classifying borderline or ambiguous content. Although a subset of posts was independently annotated and resolved through discussion, a substantial portion of the dataset was annotated only by a single Junior fact-checker, making it necessary to account for potential systematic differences in labeling.

\begin{figure}[!ht]
   \centering
    \includegraphics[width=\columnwidth]{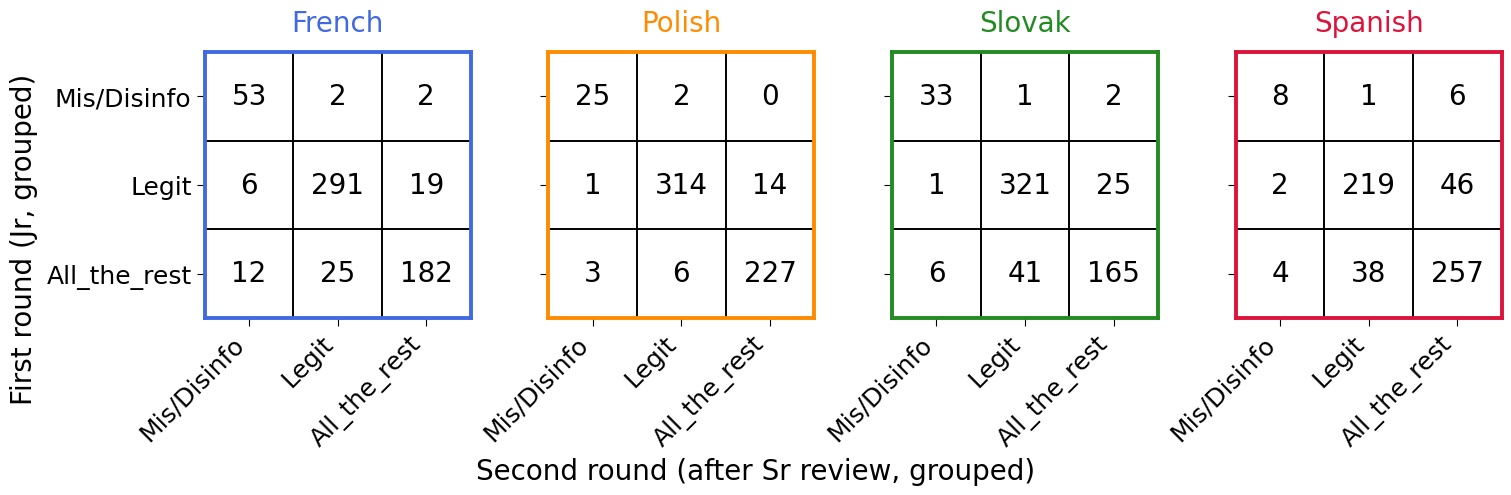}
    \caption{Junior-to-post-review label transitions in the double-coded subset, shown separately for each language, after grouping the original annotation scheme into three disjoint groups. Rows correspond to Junior first-round group assignments, and columns correspond to second-round grouped labels after Senior review.}
    \label{fig:Jr_matrix_colors}
\end{figure}

To characterise annotation errors, we analyse label transitions observed in the double-coded subsets. Figure~\ref{fig:Jr_matrix_colors} presents annotation transition matrices separately for each language, with labels grouped into three categories: \textit{Mis/Disinfo} (posts labelled as \textit{Mis/disinformation}) 
\textit{Legit} (posts labelled as \textit{Credible and informative} or \textit{Unverifiable}), corresponding to the denominator used for prevalence estimation in Equation~\ref{eq:1}); and \textit{All\_the\_rest} (all remaining labels). Rows correspond to Junior fact-checkers’ first-round grouped labels, while columns correspond to the grouped labels agreed upon after Senior review. Each cell, therefore, indicates the number of posts that transition from a given Junior category to a final post-review category.

These transition matrices provide a concise summary of annotation errors made by Junior fact-checker, including false positives and false negatives, and form the basis for correcting Junior-only annotations.

\paragraph{Multinomial Simulation of Annotation Error}

To propagate annotation-related uncertainty from the double-coded subset to the Junior-only remainder, we apply a multinomial simulation approach. This approach considers all possible annotation outcomes across the three grouped categories, rather than modelling only false-positive and false-negative rates for \textit{Mis/Disinfo}. 

For each aggregation level (language, platform, and platform–language), a separate $3\times3$ reference matrix is estimated from the corresponding double-coded subset. Each matrix represents the empirical joint distribution of Junior first-round assignments and final labels agreed upon after Senior review, with rows corresponding to Junior labels and columns corresponding to post-review labels as described in the previous section. Using these reference matrices, corrected label assignments for the Junior-only remainder are generated via repeated multinomial simulation. We performed $S=500$ simulation runs for each aggregation unit. In each simulation run, a complete $3\times3$ matrix of initial versus corrected label assignments is produced, with the total number of entries fixed to the size of the Junior-only remainder for the corresponding aggregation unit. The simulated number of posts assigned to Mis/Disinformation after correction is obtained by summing the relevant column of the simulated table and adding this value to the observed count of Mis/Disinformation posts in the corresponding double-coded subset.

\begin{figure}
    \centering
    \includegraphics[width=\columnwidth]{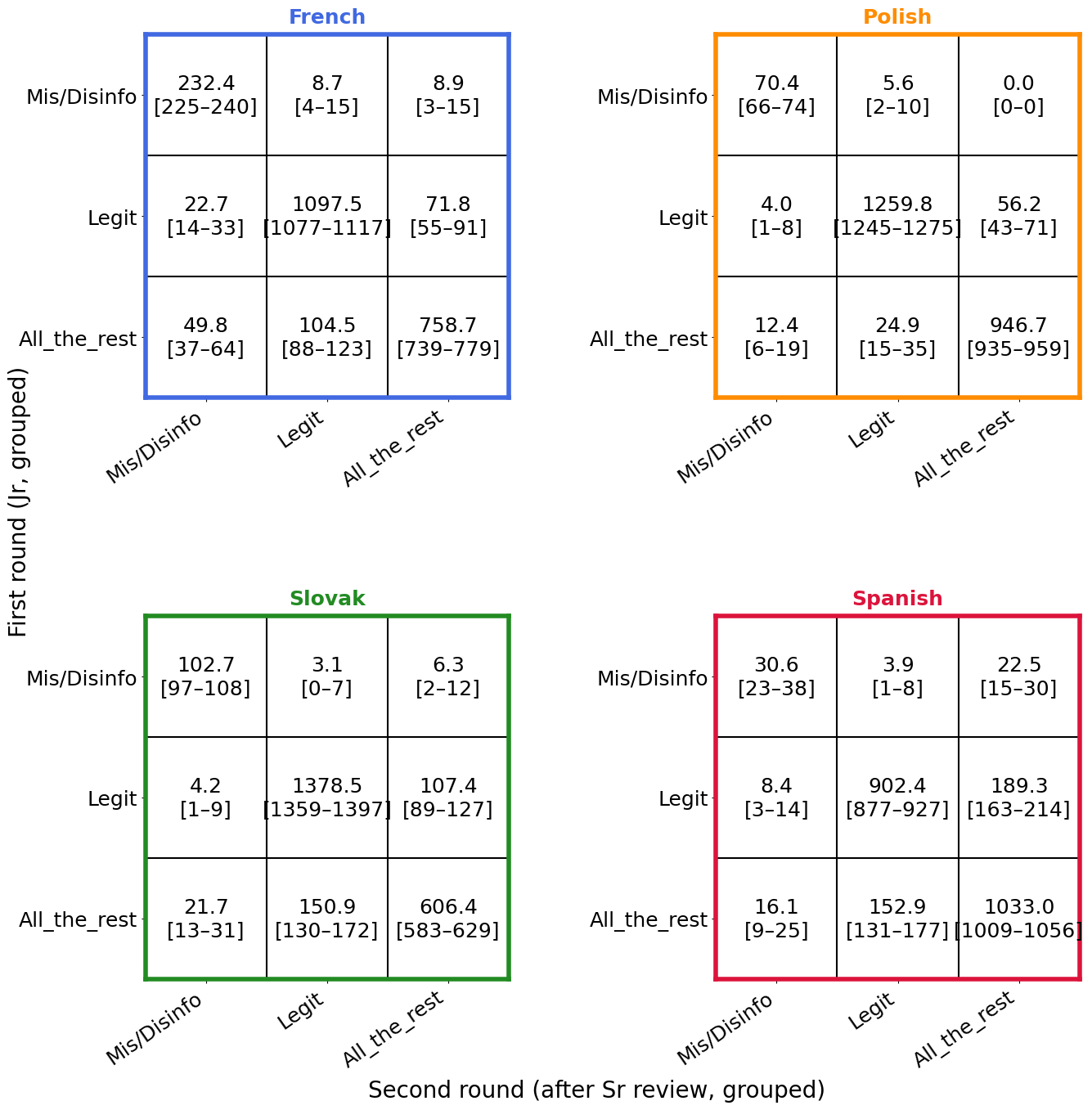}
    \caption{Mean 3×3 correction matrices under multinomial annotation uncertainty (Junior-only remainder). Each cell reports the mean expected number of posts reassigned across 500 multinomial simulations, with 2.5–97.5 percentile intervals shown in brackets.}
    \label{fig:simulation_example}
\end{figure}

Figure~\ref{fig:simulation_example} shows $3\times3$ mean correction matrices under multinomial annotation uncertainty for each language. Each panel shows the expected redistribution of Junior-only annotations across the three groups (Mis/Disinformation, Legit, and All the Rest) after applying the language-specific reference matrix estimated from double-coded posts. Algorithm~\ref{alg:multinomial} summarises the multinomial simulation procedure.

In addition to language-level prevalence estimates, we computed prevalence at both the platform level and the platform–language level to provide a more comprehensive overview. To derive prevalence estimates, we have also constructed platform and platform-language specific 3×3 reference matrices from the double-coded subset, where rows correspond to Junior annotators’ first-round labels and columns correspond to the agreed labels after Senior review (see Appendix~\ref{secA1} and Appendix~\ref{app-B}).

\begin{algorithm}
\caption{Multinomial simulation for the uncertainty derived from Junior annotation errors}
\label{alg:multinomial}
\begin{algorithmic}[1]
\renewcommand{\algorithmicrequire}{\textbf{Input:}}
\renewcommand{\algorithmicensure}{\textbf{Output:}}
\Require Dataset $A_u$ (posts in unit $u$) with Junior and agreed labels (when available), simulations $S$,  $P \leftarrow \mathrm{empty}(S)$
\Ensure Mean prevalence $\bar{\mathcal{P}}$ and 95\% CI

\State $m \leftarrow |L|$ \Comment{$L$: \{\textit{Mis/DisInfo}, \textit{Legit}, \textit{All\_the\_rest}\}}
\State $D \leftarrow$ double-coded posts in $A_u$ 

\State $R \leftarrow$ posts in $A_u$ with Junior label only

\State Build reference matrix $M \in \mathbb{R}^{m\times m}$ from $D$ (normalized rows)

\State Extract observed agreed-label counts $N_D \in \mathbb{N}^m$ from $D$
\State Extract Junior-label counts $N_R \in \mathbb{N}^m$ from $R$

\For{$s = 1$ to $S$} 
    \State Initialise simulated matrix $T \leftarrow \mathbf{0}_{m\times m}$

        \State $T \leftarrow \mathrm{Multinomial}(N_R, M)$ 
 
    \State $N^{(s)}_{1}, N^{(s)}_{2}, N^{(s)}_{3} \leftarrow N_D + \mathrm{colsum}(T)$  \Comment{Double-coded \& simulated}
    \State $P[s] \leftarrow N^{(s)}_{1} / (N^{(s)}_{1} + N^{(s)}_{2})$ \Comment{Compute prevalence}
 
\EndFor
\State $\bar{\mathcal{P}} \leftarrow \frac{1}{S} \sum_{s=1}^{S} P[s]$ \Comment{Calculate mean simulated prevalence}

\State \Return mean prevalence $\bar{\mathcal{P}}$ and 95\% CI

\end{algorithmic}
\end{algorithm}

\subsubsection{Data retrieval uncertainty}

Data retrieval uncertainty arises from the keyword-based strategy used to collect candidate posts. Although Keyword lists were generated using expert knowledge, alternative keyword selections or matches can lead to different subsets of posts being retrieved and annotated. This introduces variability in prevalence estimates.

Due to different retrieval strategies across platforms, in our corpus, keyword lists were available only for a subset of platforms, namely TikTok, YouTube, and X/Twitter. Some posts were associated with a single matched keyword. Posts without a matched keyword are assigned a dedicated \textit{empty keyword}, which is treated identically during bootstrap resampling described below. This ensures that keyword-related uncertainty is modelled consistently across both keyword-matched and non-keyword content.

To quantify data retrieval uncertainty, we applied a two-level bootstrap procedure with $B_{\mathrm{kw}} = 500$ keyword resamples and $B_{\mathrm{post}} = 500$ post resamples per keyword draw. In addition to resampling keywords, posts were also resampled to account for post-sampling uncertainty. 
For a given aggregation unit, keywords were resampled with replacement from the corresponding keyword pool, followed by resampling posts associated with each selected keyword. Prevalence was computed for each bootstrap replicate, yielding a distribution of prevalence estimates that reflects variability due to keyword choice and post-sampling (see Algorithm~\ref{alg:kw_post_bootstrap}).

\begin{algorithm}[h]
\caption{Bootstrapping for data retrieval uncertainty}
\label{alg:kw_post_bootstrap}
\begin{algorithmic}[1]
\renewcommand{\algorithmicrequire}{\textbf{Input:}}
\renewcommand{\algorithmicensure}{\textbf{Output:}}

\Require Dataset $A_u$ (posts in unit $u$) with keywords, Junior and agreed labels (when available), bootstrap sizes $B_{\mathrm{kw}}$, $B_{\mathrm{post}}$ 

\Ensure Mean prevalence $(\mathcal{P})$ and 95\% CI

\State $A \leftarrow$ unique keywords in $A_u$ 
\State $K \leftarrow |A|$ 

\State Initialise empty list $\mathcal{P} \leftarrow [\ ]$

\For{$b = 1$ to $B_{\mathrm{kw}}$} \Comment{Keyword bootstrap}
    \State $P_b \leftarrow [\ ]$
    \While {$|P_b| = 0$} \Comment{To ensure the keywords retrieved posts}
        \State Draw keywords pool $A_b$ by sampling $K$ keywords from $A$ with replacement
        \State Compute keyword multiplicities $m_{k,b}$ for each $k \in A$
        \State Construct post pool $P_b$ 
    \EndWhile    
    \For{$t = 1$ to $B_{\mathrm{post}}$} \Comment{Post bootstrap}
        \State Draw bootstrap sample $S_{b,t}$  from $P_b$ with replacement
        \State Compute prevalence $p_{b,t}$ from $S_{b,t}$
        \State Append $p_{b,t}$ to $\mathcal{P}$
        
    \EndFor
\EndFor
\State Report mean$(\mathcal{P})$ and 95\% CI 
\end{algorithmic}
\end{algorithm}

\subsubsection{Joint Uncertainty Estimation}
\paragraph{Modelling Joint Uncertainty}
The joint uncertainty estimation combines annotation-related and data retrieval uncertainty within a single estimation procedure. Rather than treating these sources independently, the approach propagates both sources of variability simultaneously to capture their combined impact on mis/disinformation prevalence estimates. 

For each bootstrap replicate, keyword-based resampling is first performed to generate a resampled set of posts. Conditional on this resampled dataset, annotation-related uncertainty is then propagated via multinomial simulation, using the corresponding reference matrix to generate corrected label assignments. Prevalence is computed for each replicate based on the simulated, corrected labels. Repeating this procedure creates a distribution of prevalence estimates that jointly reflects uncertainty arising from both keyword selection and annotation behaviour. Algorithm~\ref{alg:joint_kw_multinomial} summarises the joint uncertainty estimation procedure. 

\paragraph{Simulation Parameters}

It should be noted that the joint uncertainty procedure increases rapidly with the number of keyword bootstraps, post-bootstraps, and annotation simulations. While the previous two analyses used up to 500 iterations, the joint analysis was run with $B_{\mathrm{kw}} = 100$, $B_{\mathrm{post}} = 100$, and $S = 100$ multinomial simulations, resulting in one million simulations per language. Pilot runs indicated that both prevalence estimates and confidence intervals stabilised at this level, whereas larger numbers of iterations would have been computationally expensive.

\begin{algorithm}[t]
\caption{Joint keyword and annotation uncertainty}
\label{alg:joint_kw_multinomial}
\begin{algorithmic}[1]
\renewcommand{\algorithmicrequire}{\textbf{Input:}}
\renewcommand{\algorithmicensure}{\textbf{Output:}}

\Require Dataset $A_u$ (posts in unit $u$) with keywords, Junior and agreed labels (when available), counts $S$, $B_{\mathrm{kw}}$, $B_{\mathrm{post}}$ 

\Ensure Mean prevalence $(\mathcal{P})$ and 95\% CI

\State $A \leftarrow$ unique keywords in $A_u$ 
\State $K \leftarrow |A|$ 
\State Build the $3 \times 3$ reference matrix $M$ from double-coded posts in $A_u$

\State Initialise three-dimensional array $\mathcal{P}$ of size 
$B_{\mathrm{kw}} \times B_{\mathrm{post}} \times S$



    \For{$b = 1$ to $B_{\mathrm{kw}}$} \Comment{Keyword bootstrap}
    
    \State $P_b \leftarrow [\ ]$
    \While {$|P_b| = 0$} \Comment{To ensure the keywords retrieved posts}
        
    
      \State Draw $K$ keywords from $A$ with replacement
      
        \State Compute keyword multiplicities $m_{k,b}$
        
        \State Construct post pool $P_b$

        \EndWhile
        \For{$t = 1$ to $B_{\mathrm{post}}$} \Comment{Post sampling bootstrap}
        \State Draw bootstrap sample $S_{b,t}$ from $P_b$ with replacement

            \For{$s = 1$ to $S$} \Comment{Annotation simulation}
            
                \State Split $S_{b,t}$ into $D$ (double-coded posts) and $R$ (Junior-only posts)
                 \State Apply multinomial simulation to $R$ using reference matrix $M$
                 \State Combine simulated counts from $R$ with observed counts from $D$
            
                \State Compute prevalence and store it in $\mathcal{P}[b,t,s]$
            \EndFor
        \EndFor
    \EndFor

    \State Vectorise $\mathcal{P}$ into $\mathbf{p}$
     \State Report mean$(\mathbf{p})$ and 95\% CI

\end{algorithmic}
\end{algorithm}

\section{Results}\label{sec12}

In this section, we present the prevalence estimate $P_{\text{restricted}}$ at three levels of aggregation: by language, by platform, and by platform–language combination. Results are presented under different uncertainty assumptions, namely Baseline sample uncertainty, annotation-related uncertainty, data retrieval uncertainty, and their joint effect, to facilitate cross-unit comparison.

\subsection{Per Language Results}

\begin{figure}[H]
    \centering

    \includegraphics[width=0.98\textwidth]{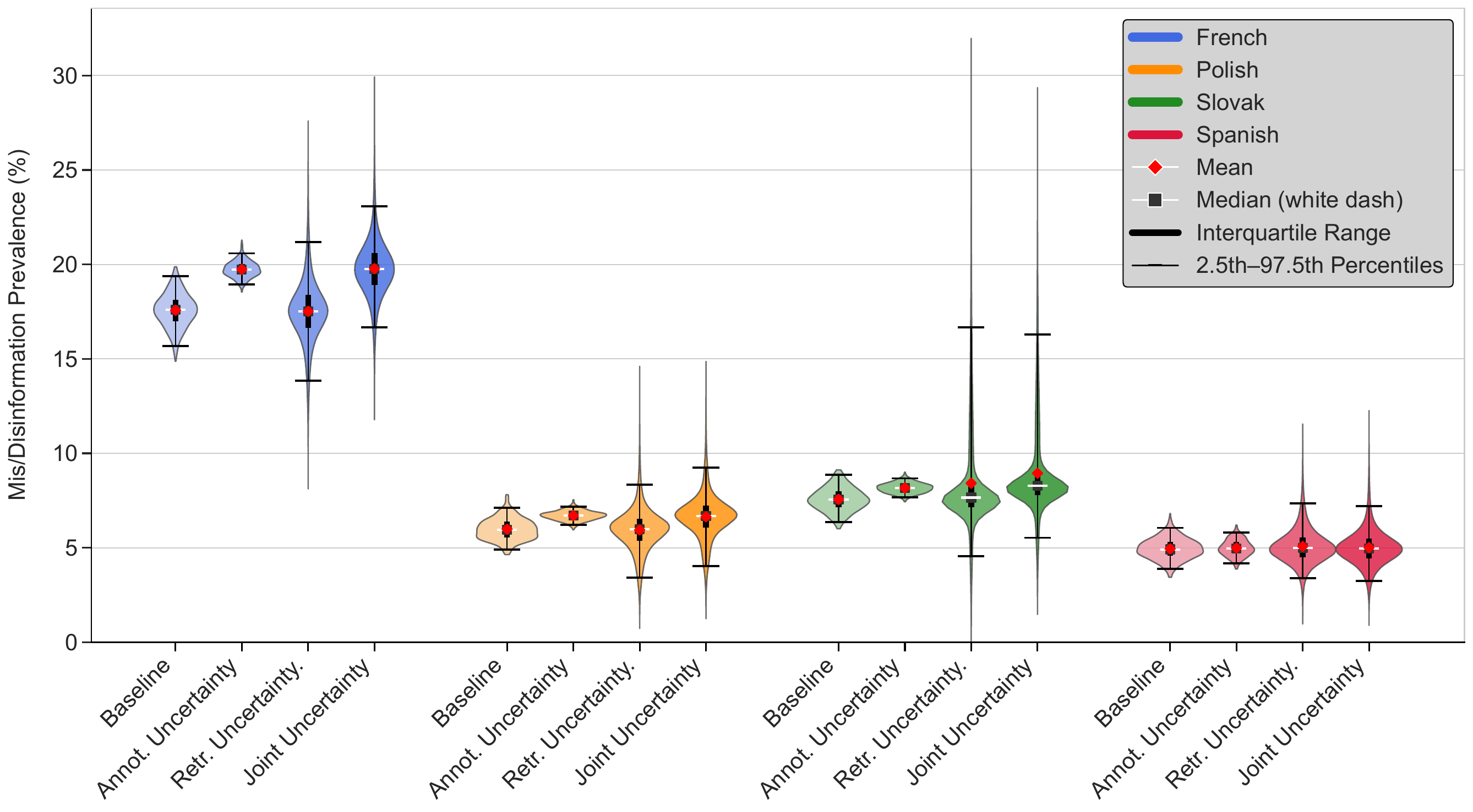}
      \caption{Violin plots of the $P_{\text{restricted}}$ estimates by language under four uncertainty assumptions. Internal markers provide a statistical summary: red diamonds for the mean, white dashes for the median, thick bars for the Interquartile Range, and capped whiskers for the 2.5th–97.5th percentiles.}
    \label{fig:per_language}
\end{figure}

Table~\ref{tab:rest_prev_by_language} and Figure~\ref{fig:per_language} summarize $P_{\text{restricted}}$ estimates by language. Baseline sample uncertainty differs substantially across languages, with French at 17.5\% [15.9–19.3], followed by Slovak at 7.6\% [6.5–8.8], Polish at 6.0\% [5.0–7.2], and Spanish at 5.0\% [4.0–6.2].

When accounting only for annotation uncertainty, prevalence estimates increase slightly for French (19.7\% [18.9–20.6]), Polish (6.7\% [6.2–7.2]), and Slovak (8.2\% [7.7–8.7]) while remaining nearly unchanged for Spanish (5.0\% [4.2–5.8]). The confidence intervals under annotation uncertainty remain narrow and comparable to the baseline intervals, suggesting that annotation-related variability contributes limited additional uncertainty beyond sampling error. For example, in French, the interval width changes only marginally despite the upward shift in the point estimate (+2.2 percentage points), indicating that annotation-related variability mainly shifts the mean but contributes limited additional dispersion beyond sample uncertainty.

In contrast, data retrieval uncertainty leads to a widening of confidence intervals across all languages. For French, the interval expands from [15.9–19.3] to [13.9–21.2]. The effect is particularly strong for Slovak, where the interval widens dramatically from [6.5–8.8] to [4.5–16.7], more than doubling its range. Polish also shows notable expansion from [5.0–7.2] to [3.4–8.3], and even Spanish—despite its relatively low baseline prevalence—widens from [4.0–6.2] to [3.4–7.3]. These results indicate that uncertainty introduced during keyword-based data collection dominates the overall variability of prevalence estimates.

Estimates under joint uncertainty exhibit the widest confidence intervals overall, with point estimates higher than baseline across all languages except Spanish. French increases to 19.8\% [16.7–23.1], Polish to 6.6\% [4.0–9.2], Slovak to 8.9\% [5.5–16.3]. However, in most cases, the joint intervals are only marginally wider than those obtained under data retrieval uncertainty alone (e.g., Slovak [4.5–16.7] vs. [5.5–16.3]). These results show that while annotation uncertainty can shift the estimate upward, retrieval-related decisions remain the primary cause of interval width and overall dispersion. 

\begin{table}[htbp]
\centering
\caption{Mis/Disinformation prevalence ($P_{\text{restricted}}$, in \%) by language under different uncertainty assumptions.}
\label{tab:rest_prev_by_language}
\small
\begin{tabular}{@{}l r@{\hskip 3pt}r r@{\hskip 3pt}r r@{\hskip 3pt}r r@{\hskip 3pt}r@{}}
\toprule
& \multicolumn{2}{c}{Baseline (Wilson)} 
& \multicolumn{2}{c}{Annot. Uncert.} 
& \multicolumn{2}{c}{Retr. Uncert.} 
& \multicolumn{2}{c}{Joint Uncert.} \\
\cmidrule(lr){2-3} \cmidrule(lr){4-5} \cmidrule(lr){6-7} \cmidrule(lr){8-9}
Language 
& Est. & 95\% CI 
& Est. & 95\% CI 
& Est. & 95\% CI 
& Est. & 95\% CI \\
\midrule
French   
& 17.5 & [15.9--19.3]
& 19.7 & [18.9--20.6]
& 17.5 & [13.9--21.2]
& 19.8 & [16.7--23.1] \\

Polish   
& 6.0 & [5.0--7.2]
& 6.7 & [6.2--7.2]
& 5.9 & [3.4--8.3]
& 6.6 & [4.0--9.2] \\

Slovak   
& 7.6 & [6.5--8.8]
& 8.2 & [7.7--8.7]
& 8.4 & [4.5--16.7]
& 8.9 & [5.5--16.3] \\

Spanish  
& 5.0 & [4.0--6.2]
& 5.0 & [4.2--5.8]
& 5.1 & [3.4--7.3]
& 5.0 & [3.2--7.2] \\
\bottomrule
\end{tabular}
\end{table}

\subsection{Per Platform Results}

\begin{figure}[h]
    \centering
   \includegraphics[width=\columnwidth]{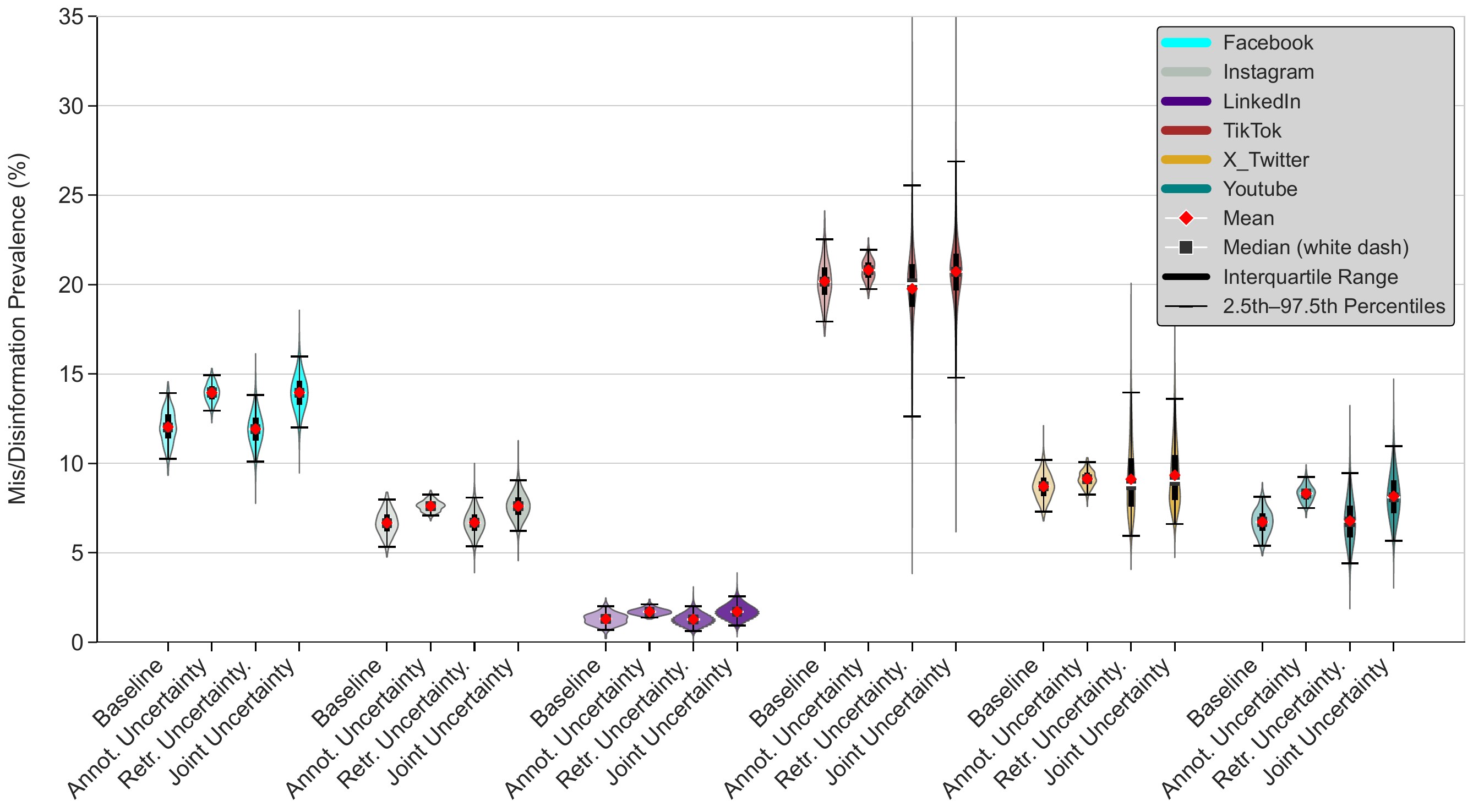}
      \caption{Mis/disinformation prevalence estimates by platform under four uncertainty assumptions. Internal markers provide a statistical summary: red diamonds for the mean, white dashes for the median, thick bars for the Interquartile Range, and capped whiskers for the 2.5th–97.5th percentiles.}
    \label{fig:per_platform}
\end{figure}

Table~\ref{tab:by_platform} and Figure~\ref{fig:per_platform} report $P_{\text{restricted}}$ estimates by platform. Baseline estimates vary across platforms, with TikTok exhibiting the highest prevalence at 20.1\% [17.9–22.6], followed by Facebook at 11.9\% [10.2–13.9] and X/Twitter at 8.7\% [7.3–10.4]. Instagram (6.7\% [5.4–8.2]) and YouTube (6.8\% [5.4–8.4]) show comparatively lower prevalence. LinkedIn has the lowest baseline estimate at 1.3\% [0.7–2.2].

When annotation uncertainty is introduced, point estimates increase across all platforms. Facebook rises from 11.9\% to 13.0\%, Instagram from 6.7\% to 7.6\%, LinkedIn from 1.3\% to 1.7\%, TikTok from 20.1\% to 20.8\%, X/Twitter from 8.7\% to 9.1\%, and YouTube shows a larger absolute increase from 6.8\% to 8.3\%. Despite these upward shifts, confidence intervals remain relatively narrow and close to the baseline widths. This indicates that annotation uncertainty primarily shifts the mean rather than substantially increasing variability. 

In contrast, retrieval uncertainty leads to a widening of confidence intervals. This effect is particularly strong for TikTok, where the interval expands from [17.9--22.6] to [12.6–25.5], X/Twitter from [7.3–10.4] to [5.9–14.0], and YouTube from [5.4–8.4] to [4.4–9.5], where confidence intervals expand substantially compared to both baseline and annotation-only estimates. Similar to the per-language analysis, these results indicate that variability introduced by keyword-based data retrieval plays a dominant role in shaping uncertainty also at the platform level.

Estimates under joint uncertainty combine annotation and retrieval effects. Also here, similar to the language-level analysis, joint uncertainty intervals are often only marginally wider than those obtained under retrieval uncertainty alone (e.g., TikTok [12.6–25.5] vs. [14.8–26.9]). This suggests that retrieval-related variability accounts for most of the total uncertainty. Overall, the platform-level results reinforce the conclusion that uncertainty in misinformation prevalence estimates is driven primarily by data retrieval choices rather than annotation uncertainty.

\begin{table}[htbp]
\centering
\caption{Mis/Disinformation prevalence ($P_{\text{restricted}}$, in \%) by platform under different uncertainty assumptions.}
\label{tab:by_platform}
\small
\begin{tabular}{@{}l r@{\hskip 3pt}r r@{\hskip 3pt}r r@{\hskip 3pt}r r@{\hskip 3pt}r@{}}
\toprule
& \multicolumn{2}{c}{Baseline (Wilson)} 
& \multicolumn{2}{c}{Annot. Uncert.} 
& \multicolumn{2}{c}{Retr. Uncert.} 
& \multicolumn{2}{c}{Joint Uncert.} \\
\cmidrule(lr){2-3} \cmidrule(lr){4-5} \cmidrule(lr){6-7} \cmidrule(lr){8-9}
Platform 
& Est. & 95\% CI 
& Est. & 95\% CI 
& Est. & 95\% CI 
& Est. & 95\% CI \\
\midrule
Facebook   
& 11.9 & [10.2--13.9]
& 13.0 & [12.9--14.9]
& 11.9 & [10.1--13.8]
& 14.0 & [12.0--16.0] \\

Instagram   
& 6.7 & [5.4--8.2]
& 7.6 & [7.1--8.2]
& 6.7 & [5.4--8.1]
& 7.6 & [6.2--9.0] \\

LinkedIn   
& 1.3 & [0.7--2.2]
& 1.7 & [1.4--2.1]
& 1.3 & [0.6--2.0]
& 1.7 & [0.9--2.6] \\

TikTok  
& 20.1 & [17.9--22.6]
& 20.8 & [19.7--21.9]
& 19.8 & [12.6--25.5]
& 20.7 & [14.8--26.9] \\

X/Twitter
& 8.7 & [7.3--10.4]
& 9.1 & [8.3--10.1]
& 9.1 & [5.9--14.0]
& 9.3 & [6.6--13.6] \\

YouTube
& 6.8 & [5.4--8.4]
& 8.3 & [7.5--9.2]
& 6.8 & [4.4--9.5]
& 8.2 & [5.7--11.0] \\

\bottomrule
\end{tabular}
\end{table}

\subsection{Platform-Language Results} \label{plt-lang-resuls}

Finally, we report $P_{\text{restricted}}$ estimates at the platform-language level, where substantial heterogeneity is observed across combinations. As shown in Table~\ref{tab:restricted_prevalence_all} (see also Figure in Appendix~\ref{app-E}), prevalence varies both across platforms within the same language and across languages within the same platform. 

Under baseline uncertainty, TikTok consistently exhibits higher prevalence across languages, particularly for French 43.1 [37.5-48.9] and Slovak 18.8 [14.7–23.7]. In contrast, LinkedIn shows very low prevalence across all languages. 

\begin{table}[b!]
\centering
\caption{Mis/Disinformation prevalence ($P_{\text{restricted}}$, in \%) by platform–language combination under different uncertainty assumptions.}
\label{tab:restricted_prevalence_all}
\small
\begin{tabular}{@{}l r@{\hskip 3pt}r r@{\hskip 3pt}r r@{\hskip 3pt}r r@{\hskip 3pt}r@{}}
\toprule
& \multicolumn{2}{c}{Baseline (Wilson)} 
& \multicolumn{2}{c}{Annot. Uncert.} 
& \multicolumn{2}{c}{Retr. Uncert.} 
& \multicolumn{2}{c}{Joint Uncert.} \\
\cmidrule(lr){2-3} \cmidrule(lr){4-5} \cmidrule(lr){6-7} \cmidrule(lr){8-9}
Combination 
& Est. & 95\% CI 
& Est. & 95\% CI 
& Est. & 95\% CI 
& Est. & 95\% CI \\
\midrule

\multicolumn{9}{l}{\textit{French (FR)}} \\
Facebook--FR   & 16.0 & [12.5--20.3] & 21.3 & [18.7--23.9] & 16.0 & [12.2--20.1] & 21.3 & [17.1--25.8] \\
Instagram--FR  & 12.5 & [9.4--16.5]  & 14.0 & [13.0--15.3] & 12.5 & [9.1--16.1]  & 14.0 & [10.4--17.9] \\
LinkedIn--FR   & 2.0  & [0.9--4.7]   & 3.7  & [2.3--5.3]   & 2.0  & [0.4--4.0]   & 3.7  & [1.6--6.2] \\
TikTok--FR     & 43.1 & [37.5--48.9] & 42.3 & [40.3--44.2] & 41.4 & [23.0--52.9] & 40.1 & [21.7--52.1] \\
X/Twitter--FR  & 20.0 & [15.9--24.8] & 18.1 & [15.4--20.8] & 20.1 & [12.1--29.9] & 18.3 & [12.3--25.5] \\
YouTube--FR    & 10.9 & [7.9--14.7]  & 15.0 & [12.9--17.0] & 10.7 & [5.0--17.0]  & 15.1 & [9.5--20.9] \\
\midrule

\multicolumn{9}{l}{\textit{Polish (PL)}} \\
Facebook--PL   & 9.0  & [6.3--12.7]  & 9.0  & [9.0--9.0]   & 9.0  & [6.0--12.3]  & 9.0  & [5.9--12.4] \\
Instagram--PL  & 9.7  & [6.9--13.6]  & 11.4 & [10.2--12.7] & 9.7  & [6.5--13.2]  & 11.4 & [7.9--15.2] \\
LinkedIn--PL   & 0.0  & [0.0--1.1]   & 0.0  & [0.0--0.0]   & 0.0  & [0.0--0.0]   & 0.0  & [0.0--0.0] \\
TikTok--PL     & 7.6  & [4.8--11.9]  & 10.1 & [8.3--11.8]  & 7.7  & [0.0--19.2]  & 10.6 & [2.6--22.9] \\
X/Twitter--PL  & 6.1  & [3.9--9.3]   & 6.3  & [5.0--7.9]   & 6.6  & [2.6--14.3]  & 7.1  & [3.5--12.7] \\
YouTube--PL    & 4.4  & [2.6--7.6]   & 5.2  & [3.7--7.1]   & 4.5  & [1.2--8.6]   & 5.2  & [2.2--8.8] \\
\midrule

\multicolumn{9}{l}{\textit{Slovak (SK)}} \\
Facebook--SK   & 14.6 & [11.3--18.7] & 17.5 & [16.0--19.3] & 14.6 & [11.0--18.4] & 17.5 & [13.8--21.4] \\
Instagram--SK  & 0.7  & [0.2--2.1]   & 0.7  & [0.7--0.7]   & 0.7  & [0.0--1.7]   & 0.7  & [0.0--1.6] \\
LinkedIn--SK   & 1.8  & [0.8--4.2]   & 1.8  & [1.7--1.9]   & 1.8  & [0.4--3.5]   & 1.8  & [0.4--3.5] \\
TikTok--SK     & 18.8 & [14.7--23.7] & 20.4 & [18.0--23.1] & 17.6 & [4.9--26.5]  & 19.5 & [8.7--28.3] \\
X/Twitter--SK  & 4.7  & [3.1--7.0]   & 4.7  & [4.6--4.8]   & 4.7  & [2.8--6.7]   & 4.7  & [2.9--6.8] \\
YouTube--SK    & 7.5  & [4.7--11.7]  & 7.4  & [5.7--9.2]   & 7.6  & [2.7--14.6]  & 7.8  & [3.2--14.2] \\
\midrule

\multicolumn{9}{l}{\textit{Spanish (ES)}} \\
Facebook--ES   & 3.5  & [1.6--7.5]   & 1.5  & [0.6--2.9]   & 3.5  & [1.1--6.5]   & 1.6  & [0.0--3.7] \\
Instagram--ES  & 4.9  & [2.8--8.4]   & 7.9  & [6.0--10.4]  & 4.9  & [2.4--7.8]   & 7.9  & [4.8--11.2] \\
LinkedIn--ES   & 1.8  & [0.6--5.3]   & 2.2  & [2.1--2.3]   & 1.8  & [0.0--4.2]   & 2.2  & [0.0--5.1] \\
TikTok--ES     & 8.0  & [5.4--11.7]  & 7.6  & [5.3--9.9]   & 7.8  & [0.0--15.9]  & 8.5  & [2.0--19.7] \\
X/Twitter--ES  & 5.8  & [3.6--9.2]   & 6.3  & [4.3--8.6]   & 6.0  & [2.4--10.8]  & 6.5  & [3.2--10.7] \\
YouTube--ES    & 3.8  & [2.1--6.6]   & 4.5  & [4.2--4.7]   & 3.8  & [1.1--7.3]   & 4.9  & [1.5--9.2] \\

\bottomrule
\end{tabular}
\end{table}

Accounting for annotation uncertainty via multinomial simulation leads to upward shifts in point estimates for several platform-language combinations (e.g., Facebook-French from 16\% to 21.3\%, YouTube-French from 10.9\% to 15.0\%, TikTok-Polish from 7.6\% to 10.1\%). These shifts indicate the correction of annotation bias in specific platform-language contexts. However, confidence intervals remain comparatively narrow (e.g., TikTok-French from [37.5–48.9] to [40.3–44.2], showing that annotation uncertainty mainly influences the central estimate rather than the variability. In several LinkedIn combinations (e.g., LinkedIn-Polish), prevalence is 0.0\% under all assumptions, reflecting sparse misinformation cases. 

In contrast, retrieval uncertainty produces increased confidence intervals across nearly all combinations. This effect is particularly strong for TikTok across all languages, where intervals widen substantially and become highly asymmetric, reflecting strong sensitivity to keyword-based sampling variability (e.g., TikTok–French expands from [37.5–48.9] to [23.0–52.9]). Retrieval uncertainty also increases confidence intervals for X/Twitter and YouTube, even when baseline prevalence is moderate. These results indicate that variability introduced during data retrieval dominates uncertainty at the platform–language level. As expected, retrieval uncertainty shows only minor or no changes for platforms where content was not retrieved via keyword-based filtering, i.e., Facebook (e.g., Facebook–French [12.5–20.3] vs. [12.2–20.1]), Instagram, and LinkedIn.

Estimates under joint uncertainty show the widest confidence intervals overall, but in most cases, they are only marginally wider than those obtained under retrieval uncertainty alone (e.g., TikTok–French moves from [23.0–-52.9] (retrieval) to [21.7–52.1] (joint)). This pattern holds consistently across languages and platforms, suggesting that once retrieval uncertainty is accounted for, annotation uncertainty adds comparatively little additional dispersion. Nevertheless, annotation correction still shifts the point estimate itself rather than substantially increasing the variability of the estimate, accounting for a potential bias of the junior annotator.

\section{Discussion}\label{discussion}

Obtaining reliable estimates of mis/disinformation prevalence in social media is essential for informing policy, platform governance, and public debate, yet such estimation remains methodologically challenging. Recent reviews highlight substantial heterogeneity in misinformation research, including differences in annotation procedures, sampling strategies, data collection methods, and analytical techniques \cite{suarez2021prevalence}. Such variation complicates direct comparison of prevalence estimates across studies~\cite{kbaier2024prevalence}. Moreover, much of the existing research focuses on single platforms or individual countries, limiting cross-context comparability. Other recent initiatives have also underscored the methodological challenges of constructing reliable prevalence estimates when operating with limited-scale datasets \cite{trustlab2023pilot}.

Different from this background, our study advances the field by extracting a comparatively large, multi-country, multi-platform corpus and by systematically quantifying both annotation and retrieval uncertainties, as well as their combined impact on prevalence estimation. Specifically, we quantify three major sources of uncertainty, namely annotation uncertainty and data retrieval uncertainty, and also assess their joint effects to examine the robustness of prevalence estimates under different methodological conditions.

Based on our study findings, data retrieval uncertainty (estimated through a bootstrap procedure) contributed substantially to the uncertainty and widened the confidence intervals. Data retrieval uncertainty captures the variability in prevalence estimates induced by the choice of keywords used to retrieve content. Because keyword lists provide a partial representation of the underlying mis/disinformation space, different keyword selections can lead to different observed samples and, consequently, different prevalence estimates \cite{suarez2021prevalence, trustlab2023pilot}. This was observed mainly for Polish and Slovak at the language level and related to TikTok at the platform-language level.

Annotation uncertainty, modelled through multinomial simulation, leads to moderate adjustments in prevalence estimates primarily through shifts in the point estimates. In particular, correcting annotation errors can move the estimate upward or downward, reflecting annotation bias. However, these adjustments tend to affect the central estimate more than the overall width of the confidence intervals. The largest uncertainties are observed under joint uncertainty, where both data retrieval and annotation uncertainties are considered. This indicates that prevalence estimates are sensitive to multiple sources of uncertainty. These findings highlight the importance of explicitly quantifying and reporting uncertainty estimates alongside the baseline prevalence estimates.

The results also suggest several strategies for reducing uncertainty in future mis/disinformation prevalence studies. First, regarding the data retrieval uncertainty, it can be reduced by improving how keywords are chosen when collecting data. Given that the chosen keywords determine which posts are included in the corpus, limited keyword sets can miss important parts of the mis/disinformation space or create a bias and may over-represent specific narratives. One way to reduce this uncertainty is to expand keyword lists such that they capture a wider range of mis/disinformation narratives. This means including not only general topic terms, but also synonyms, spelling variations, and commonly used hashtags. When multiple keywords describe the same idea, the results become less dependent on any single keyword. Moreover, it is also possible to balance keywords across different themes or narratives. If a small number of keywords dominate the data collection, removing or replacing one of them can strongly affect the results. Using several related keywords for each narrative helps ensure that all major themes are consistently represented.

In addition, keyword lists can also be tested empirically for their reliability. For instance, the sensitivity of the prevalence estimates based on the inclusion or exclusion of individual keywords (e.g., removing a keyword leads to large changes in estimated prevalence) may indicate whether the keywords set needs to be adjusted. These strategies may help to reduce uncertainty caused by keyword selection.

Annotation uncertainty comes from differences or mistakes in how content is labelled during the annotation process. This can lead to some mis/disinformation being missed or some non-misinformation being incorrectly labelled. Annotation-related uncertainty can be reduced by improving the reliability of the annotation process. One important aspect, as also shown in this paper, is to use double-coding, where a subset of posts is independently annotated by more than one annotator, e.g., a senior fact-checker \cite{artstein2009bias}. Comparing these annotations makes it possible to model disagreements. In addition, increasing the number of annotators may also help to reduce annotation-related uncertainty by improving label reliability. 

However, our results indicate that annotation uncertainty contributes less to the overall variability of the estimate, as retrieval uncertainty contributes more to widening the confidence interval. However, annotation correction still produces noticeable shifts in the point estimate in several platform-language combinations, reflecting annotation bias. This suggests that modelling annotator error remains important for improving estimate accuracy, even though it contributes less to overall variability.

\section{Conclusions}\label{conclusions}

The spread of mis/disinformation in social media undermines trust in online information \cite{ozcelik2025detecting}, influences public opinion, and poses risks to societal well-being \cite{rashid2025probabilistic}. Consequently, detecting and monitoring mis/disinformation has become a central challenge. An important first step in addressing this challenge is to produce reliable estimates of its prevalence on social media. Large-scale benchmarking initiatives under the EU Code of Practice have demonstrated both the feasibility and the methodological complexity of measuring prevalence across platforms \cite{trustlab2023pilot}. In line with these efforts, the present study estimated mis/disinformation prevalence while explicitly modelling multiple sources of uncertainty. The results indicate that data retrieval decisions, particularly keyword selection, have a stronger influence on prevalence estimates than annotation reliability. These findings underscore the importance of reporting not only baseline prevalence estimates but also the associated uncertainty measures to enable more transparent and robust interpretation. 

Several directions for future research emerge from this analysis. First, improving keyword selection strategies is essential. Our findings show that prevalence estimates are highly sensitive to the keywords used for data collection. Future studies should systematically assess how individual keywords or keyword groups influence prevalence estimates, identifying which contribute to increased uncertainty and which produce stable results. Such analyses would support robust and transparent retrieval strategies.

In addition, social media mis/disinformation research is increasingly shaped by structural constraints in platform data access. As noted in prior work \cite{suarez2021prevalence} and recent benchmarking efforts under the EU Code of Practice \cite{trustlab2023pilot}, restricted data availability, reduced API access, and regulatory changes limit the scope and comparability of prevalence measurement. These constraints affect which platforms and content types can be studied. It may introduce selection bias and incomplete coverage. Future research should therefore explore standardized and transparent frameworks for data access and documentation to enhance comparability across studies. Moreover, annotation-related uncertainty can be addressed by strengthening the reliability of the labeling process. Clearer annotation guidelines and training may improve annotator agreement levels. Hybrid approaches that combine expert human annotation with AI-assisted review may further enhance consistency when working with large datasets.

\subsection{Policy Implications}
Several policy implications can be extracted from our findings. Prevalence estimates should not be reported without accompanying uncertainty measures, as estimates alone may create a misleading sense of precision. Reporting confidence intervals and documenting sources of uncertainty should be required to become standard practice in monitoring frameworks. Moreover, stable and timely access to platform data is essential for research. During this study, difficulties in obtaining data, regarding general access as well as keyword-specific retrieval, introduced delays and contributed to retrieval-related uncertainty. Regulatory frameworks, therefore, need to ensure effective data access mechanisms, including clear and time-efficient procedures to review and resolve cases in which data requests are denied or delayed. Strengthening such mechanisms would improve accountability and support more robust and comparable prevalence monitoring efforts.

\subsection{Limitations and Future Work}
This study has several limitations. First, considering the data collected, the dataset used captures mis/disinformation prevalence at a specific period. However, mis/disinformation narratives evolve over time, and uncertainty patterns may change accordingly \cite{kbaier2024prevalence}. Moreover, the study has considered four languages and six platforms. Extending the analysis to additional languages, platforms, or time periods would allow assessment of how uncertainty evolves across contexts and over time. We have modelled uncertainty arising from two main sources: data retrieval and annotators. Future research could extend this work by modelling additional sources of uncertainty that were not considered in this study to further improve the robustness of mis/disinformation prevalence estimates. For instance, modelling domain-specific variation (e.g., health, politics, climate, science, and technology domains) in annotation behaviour may offer a more fine-grained understanding of how uncertainty shapes prevalence results across content domains.

\newpage
\backmatter

\section*{Declarations}

\bmhead{Funding}
This work has been funded by the European Media and Information Fund. Ref: 101177191. 

\bmhead{Acknowledgements}
We would like to thank the partners of the SIMODS project. Special thanks go to Daiana Crisan for facilitating the data collection and sharing. 

\bmhead{Ethics Approval}
The project received ethical approval from the Ethics Committee of Universitat Oberta Catalunya (Reference number CE24-AA56).

\bmhead{Availability of data and materials}

To ensure transparency and reproducibility, the code and the annotated dataset are publicly available at: https://tinyurl.com/simods\\
The full list of keywords is available to researchers upon request for legitimate scientific purposes (contact: data-requests@feedback.org). However, due to platform data access restrictions and compliance requirements, the mapping between specific keywords and the corresponding retrieved content cannot be publicly shared.

\bmhead{Authors’ contributions}
A.K and I.A. conceptualised the study. I. A. was responsible for writing the code, conducting the analyses, and drafting the manuscript. A.K. was responsible for overall supervision and contributed to the analyses and the drafting of the manuscript. S.R. and J.A. contributed conceptual insights. E.V. was responsible for data collection and contributed conceptual insights. All authors contributed to writing and revising the manuscript.

\bmhead{Competing interests}
The authors declare no competing interests.

\newpage
\begin{appendices} \label{app-A}

\section{Reference matrices per platform}\label{secA1}

This appendix is related to the content of Section \ref{annot_uncert} of the manuscript and presents the platform-specific reference matrices derived from the double-coded subset. The matrices summarise the transition patterns between independent Junior annotations and the final grouped labels assigned after the resolution phase.

\begin{figure}[!ht]
    \centering
    \includegraphics[width=\columnwidth]{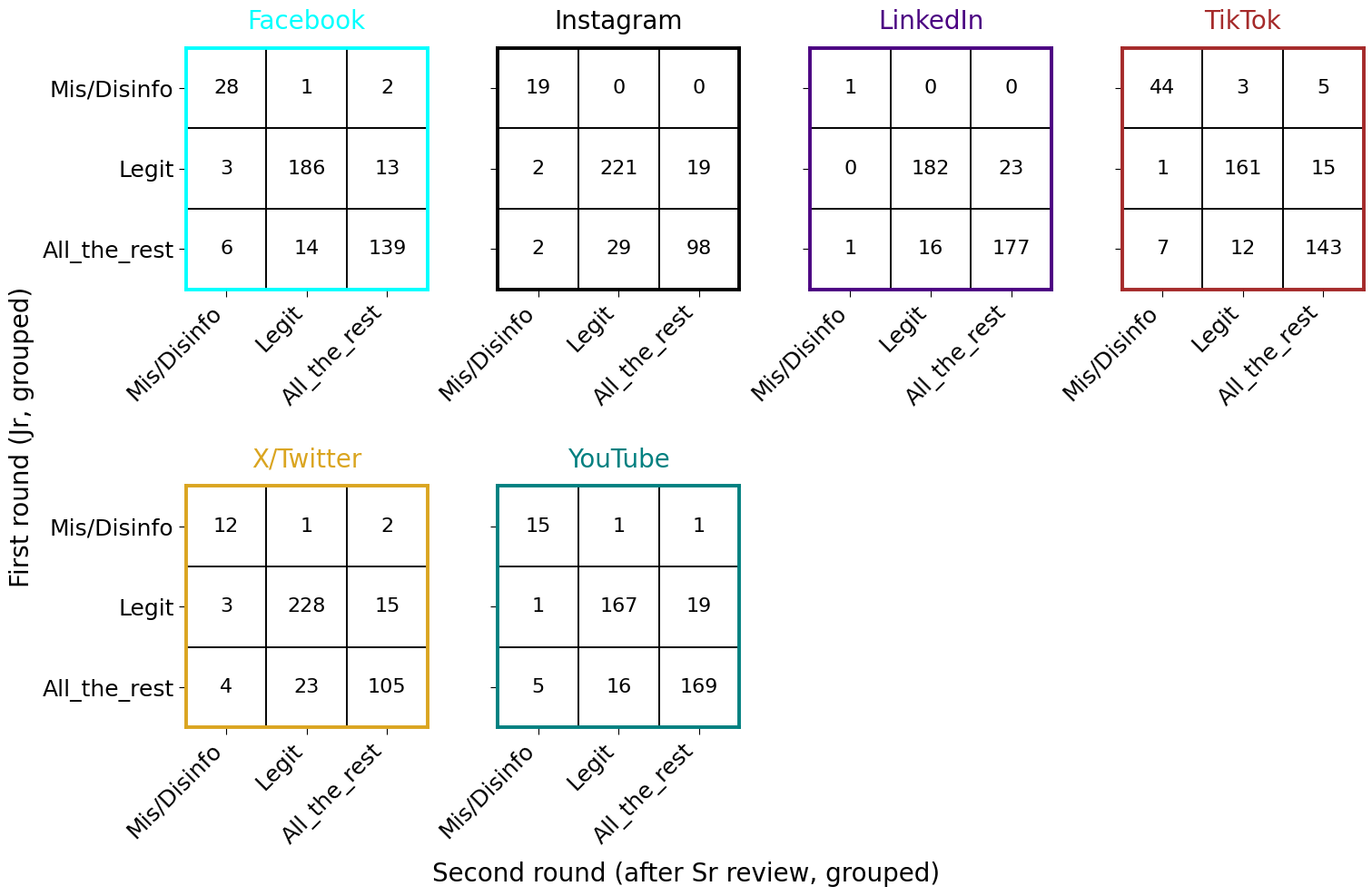}
    \caption{Junior-to-post-review label transitions in the double-coded subset, shown separately for each platform, after grouping the original annotation scheme into three disjoint groups. Rows correspond to Junior first-round group assignments, and columns correspond to second-round grouped labels after Senior review.}
    \label{fig:plat_matrix_appendix}
\end{figure}

\clearpage
\section{Reference matrices for all platform-language combinations} \label{app-B}

This appendix is related to the content of Section \ref{annot_uncert} of the manuscript and presents the reference matrices for each platform-language combination derived from the double-coded subset. These matrices capture the transition patterns between Junior independent annotations and the final grouped labels assigned after the resolution phase.

\begin{figure}[H]
    \centering
    \includegraphics[width=0.8\textwidth, height=0.7\textheight]{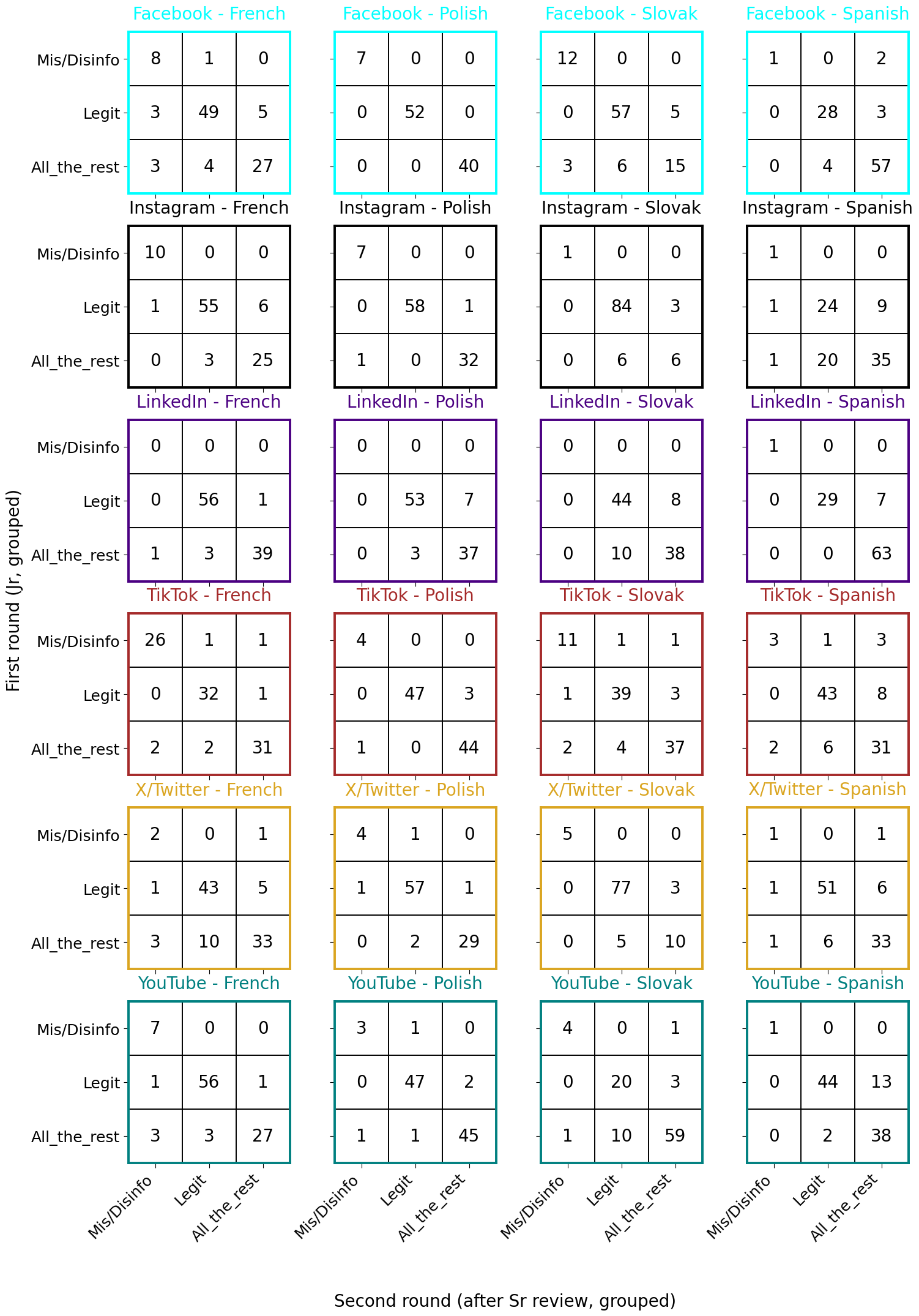}
    \caption{Junior-to-post-review label transitions in the double-coded subset. Rows correspond to Junior first-round group assignments, and columns correspond to second-round grouped labels after Senior review.}
    \label{fig:all_matrix_appendix}
\end{figure}

\newpage
\section{Complementary calculation of mis/disinformation prevalence ($P_{\text{total}}$)}   \label{app-C}

This appendix reports the complementary prevalence estimates, $P_{\text{total}}$ (see Equation~\ref{eq:2}), under different uncertainty assumptions. 

First, considering estimates per language, as shown in Table~\ref{tab:total_prevalence_language}, $P_{\text{total}}$ varies systematically across languages and uncertainty assumptions, with patterns consistent with those observed for $P_{\text{restricted}}$. Across all methods, French exhibits the highest baseline sample uncertainty, followed by Slovak, Polish, and Spanish. Under the baseline (Wilson) estimate, $P_{\text{total}}$ ranges from 2.4\% (Spanish) to 10.9\% (French). When annotation uncertainty is incorporated, prevalence increases slightly for most languages (e.g., French from 10.9\% to 12.8\%).

\begin{table}[!b]
\centering
\caption{$P_{\text{total}}$ (in \%) by language under different uncertainty assumptions.}
\label{tab:total_prevalence_language}
\small
\begin{tabular}{@{}l r@{\hskip 3pt}r r@{\hskip 3pt}r r@{\hskip 3pt}r r@{\hskip 3pt}r@{}}
\toprule
& \multicolumn{2}{c}{Baseline (Wilson)} 
& \multicolumn{2}{c}{Annot. Uncert.} 
& \multicolumn{2}{c}{Retr. Uncert.} 
& \multicolumn{2}{c}{Joint Uncert.} \\
\cmidrule(lr){2-3} \cmidrule(lr){4-5} \cmidrule(lr){6-7} \cmidrule(lr){8-9}
Language 
& Est. & 95\% CI 
& Est. & 95\% CI 
& Est. & 95\% CI 
& Est. & 95\% CI \\
\midrule
French   
& 10.9 & [9.8--12.1]
& 12.8 & [12.2--13.4]
& 11.0 & [8.8--13.7]
& 12.9 & [10.9--15.6] \\

Polish   
& 3.5 & [2.9--4.3]
& 3.9 & [3.6--4.2]
& 3.4 & [2.0--4.6]
& 3.8 & [2.2--5.0] \\

Slovak   
& 5.1 & [4.4--6.0]
& 5.7 & [5.3--6.0]
& 5.3 & [2.7--9.4]
& 5.9 & [3.5--10.0] \\

Spanish  
& 2.4 & [1.9--3.0]
& 2.4 & [1.9--2.8]
& 2.6 & [1.8--4.2]
& 2.4 & [1.7--3.7] \\

\bottomrule
\end{tabular}
\end{table}

Retrieval uncertainty produces substantially wider confidence intervals, particularly for Slovak (5.3\% [2.7–9.4]) and French (11.0\% [8.8–13.7]), indicating that keyword-based sampling variability is a major contributor to total uncertainty. This pattern is closely related to the widening of intervals in $P_{\text{restricted}}$, where keyword/post bootstrapping also introduced a higher dispersion than annotation simulation alone.

Finally, the joint uncertainty model, combining annotation and retrieval uncertainty, resulted in the widest intervals overall, confirming that both sources of uncertainty meaningfully contribute to prevalence variability. However, the ranking of languages remains stable across all specifications, suggesting that the comparative differences observed in $P_{\text{restricted}}$ are robust to alternative uncertainty assumptions.

Table~\ref{tab:total_prevalence_platforms} presents $P_{\text{total}}$ by platform under different uncertainty assumptions. The results show clear cross-platform differences yet remain consistent to the results observed for $P_{\text{restricted}}$. TikTok exhibits the highest prevalence (baseline: 11.2\%) while LinkedIn consistently reports the lowest prevalence (0.7\%). This observation remains unchanged across baseline, annotation-only, retrieval-only, and joint models, indicating robust platform-level differences.

Incorporating annotation uncertainty slightly increases point estimates across most platforms, consistent with the observations for $P_{\text{restricted}}$ when adjusting Junior-only labels through multinomial simulation.

Retrieval uncertainty primarily affects interval width. For example, TikTok’s interval expands substantially under retrieval uncertainty, e.g., [6.1–13.2], highlighting the contribution of keyword-level sampling variability. This widening of confidence intervals is also observed in $P_{\text{restricted}}$ under keyword/post bootstrapping.

\begin{table}[htbp]
\centering
\caption{$P_{\text{total}}$ (in \%) by platform under different uncertainty assumptions.}
\label{tab:total_prevalence_platforms}
\small
\begin{tabular}{@{}l r@{\hskip 3pt}r r@{\hskip 3pt}r r@{\hskip 3pt}r r@{\hskip 3pt}r@{}}
\toprule
& \multicolumn{2}{c}{Baseline (Wilson)} 
& \multicolumn{2}{c}{Annot. Uncert.} 
& \multicolumn{2}{c}{Retr. Uncert.} 
& \multicolumn{2}{c}{Joint Uncert.} \\
\cmidrule(lr){2-3} \cmidrule(lr){4-5} \cmidrule(lr){6-7} \cmidrule(lr){8-9}
Platform 
& Est. & 95\% CI 
& Est. & 95\% CI 
& Est. & 95\% CI 
& Est. & 95\% CI \\
\midrule
Facebook   
& 7.1 & [6.0--8.3]
& 8.4 & [7.8--9.0]
& 7.1 & [5.9--8.2]
& 8.4 & [7.2--9.7] \\

Instagram   
& 4.4 & [3.5--5.3]
& 5.2 & [4.9--5.6]
& 4.4 & [3.5--5.3]
& 5.2 & [4.3--6.2] \\

LinkedIn   
& 0.7 & [0.4--1.1]
& 0.9 & [0.7--1.1]
& 0.7 & [0.3--1.0]
& 0.9 & [0.5--1.3] \\

TikTok  
& 11.2 & [9.9--12.7]
& 11.7 & [11.0--12.4]
& 10.6 & [6.1--13.2]
& 11.4 & [7.5--14.2] \\

X/Twitter
& 5.9 & [5.0--7.1]
& 6.4 & [5.7--7.0]
& 6.0 & [4.2--8.4]
& 6.4 & [4.7--8.7] \\

YouTube
& 3.8 & [3.0--4.7]
& 4.6 & [4.1--5.1]
& 3.8 & [2.5--5.5]
& 4.6 & [3.2--6.2] \\

\bottomrule
\end{tabular}
\end{table}

The joint uncertainty model produces the widest intervals overall, confirming that annotation and retrieval uncertainties jointly increase the variability of the estimates. Moreover, the relative ordering of platforms remains stable, suggesting that comparative platform differences in both $P_{\text{total}}$ and $P_{\text{restricted}}$.

\begin{table}[htbp]
\centering
\caption{$P_{\text{total}}$ in (\%) by platform–language combination under different uncertainty assumptions.}
\label{tab:merged_total_platform_language}
\small
\begin{tabular}{@{}l r@{\hskip 3pt}r r@{\hskip 3pt}r r@{\hskip 3pt}r r@{\hskip 3pt}r@{}}
\toprule
& \multicolumn{2}{c}{Baseline (Wilson)} 
& \multicolumn{2}{c}{Annot. Uncert.} 
& \multicolumn{2}{c}{Retr. Uncert.} 
& \multicolumn{2}{c}{Joint Uncert.} \\
\cmidrule(lr){2-3} \cmidrule(lr){4-5} \cmidrule(lr){6-7} \cmidrule(lr){8-9}
Combination 
& Est. & 95\% CI 
& Est. & 95\% CI 
& Est. & 95\% CI 
& Est. & 95\% CI \\
\midrule

\multicolumn{9}{l}{\textit{French (FR)}} \\
Facebook--FR   & 10.6 & [8.2--13.7]  & 14.5 & [12.7--16.5] & 10.6 & [8.0--13.5]  & 14.5 & [11.4--17.7] \\
Instagram--FR  & 8.5  & [6.3--11.3]  & 9.3  & [8.7--10.1]  & 8.5  & [6.1--11.1]  & 9.3  & [6.9--11.9] \\
LinkedIn--FR   & 1.0  & [0.4--2.3]   & 2.0  & [1.2--2.8]   & 1.0  & [0.2--2.0]   & 2.0  & [0.8--3.2] \\
TikTok--FR     & 25.9 & [22.2--30.0] & 26.4 & [24.9--28.0] & 25.0 & [13.1--34.0] & 25.0 & [13.0--33.1] \\
X/Twitter--FR  & 12.5 & [9.8--15.7]  & 11.4 & [9.4--13.3]  & 12.5 & [7.4--18.8]  & 11.5 & [7.8--15.9] \\
YouTube--FR    & 7.2  & [5.2--9.8]   & 10.6 & [9.0--12.1]  & 7.1  & [3.3--11.2]  & 10.7 & [6.7--14.9] \\
\midrule

\multicolumn{9}{l}{\textit{Polish (PL)}} \\
Facebook--PL   & 5.6 & [3.9--8.0] & 5.6 & [5.6--5.6] & 5.6 & [3.6--7.6] & 5.6 & [3.6--7.6] \\
Instagram--PL  & 5.8 & [4.1--8.3] & 6.8 & [6.0--7.8] & 5.8 & [3.8--8.0] & 6.9 & [4.6--9.3] \\
LinkedIn--PL   & 0.0 & [0.0--0.8] & 0.0 & [0.0--0.0] & 0.0 & [0.0--0.0] & 0.0 & [0.0--0.0] \\
TikTok--PL     & 3.5 & [2.2--5.5] & 4.5 & [3.7--5.3] & 3.3 & [0.0--7.0] & 4.3 & [0.9--8.1] \\
X/Twitter--PL  & 3.8 & [2.5--5.9] & 4.1 & [3.2--5.1] & 4.1 & [1.6--8.4] & 4.5 & [2.2--7.8] \\
YouTube--PL    & 2.4 & [1.4--4.2] & 2.8 & [2.0--3.8] & 2.4 & [0.7--4.7] & 2.8 & [1.2--4.9] \\
\midrule

\multicolumn{9}{l}{\textit{Slovak (SK)}} \\
Facebook--SK   & 10.6 & [8.2--13.6] & 13.6 & [12.2--15.0] & 10.6 & [7.9--13.4] & 13.5 & [10.6--16.5] \\
Instagram--SK  & 0.6  & [0.2--1.8]  & 0.6  & [0.6--0.6]   & 0.6  & [0.0--1.4]  & 0.6  & [0.0--1.4] \\
LinkedIn--SK   & 1.0  & [0.4--2.3]  & 1.0  & [1.0--1.0]   & 1.0  & [0.2--2.0]  & 1.0  & [0.2--2.0] \\
TikTok--SK     & 10.9 & [8.5--14.0] & 12.1 & [10.7--13.8] & 10.0 & [2.5--14.6] & 11.3 & [4.7--16.5] \\
X/Twitter--SK  & 4.2  & [2.8--6.4]  & 4.2  & [4.2--4.2]   & 4.2  & [2.6--6.0]  & 4.2  & [2.6--6.0] \\
YouTube--SK    & 3.4  & [2.2--5.4]  & 3.5  & [2.6--4.4]   & 3.7  & [1.5--7.9]  & 3.9  & [1.7--7.9] \\
\midrule

\multicolumn{9}{l}{\textit{Spanish (ES)}} \\
Facebook--ES   & 1.2 & [0.6--2.7] & 0.5 & [0.2--1.0] & 1.2 & [0.4--2.3] & 0.6 & [0.0--1.2] \\
Instagram--ES  & 2.5 & [1.4--4.3] & 4.3 & [3.3--5.6] & 2.5 & [1.2--3.9] & 4.3 & [2.7--6.2] \\
LinkedIn--ES   & 0.6 & [0.2--1.8] & 0.6 & [0.6--0.6] & 0.6 & [0.0--1.4] & 0.6 & [0.0--1.4] \\
TikTok--ES     & 4.7 & [3.2--7.0] & 4.4 & [3.0--5.7] & 4.4 & [0.0--8.1] & 4.5 & [0.9--9.9] \\
X/Twitter--ES  & 3.2 & [2.0--5.2] & 3.5 & [2.4--5.0] & 3.3 & [1.3--6.0] & 3.7 & [1.8--6.2] \\
YouTube--ES    & 2.2 & [1.2--3.9] & 2.2 & [2.2--2.2] & 2.2 & [0.6--4.3] & 2.4 & [0.7--4.6] \\

\bottomrule
\end{tabular}
\end{table}

Table~\ref{tab:merged_total_platform_language} shows heterogeneity in $P_{\text{total}}$ across platform–language combinations.
The highest prevalence is consistently observed for TikTok–-French (25.9\%), followed by for instance, X/Twitter–-French, TikTok–-Slovak, and Facebook–Slovak, all exceeding 10\% under most specifications. In contrast, several combinations (particularly LinkedIn across languages and Instagram in Slovak) remain below 1–2\%. 

Annotation uncertainty tends to increase estimates in several platform-language combinations (e.g., Facebook–-French, Facebook–-Slovak, YouTube–-French). Retrieval uncertainty mainly widens confidence intervals, most notably for TikTok–-French, TikTok–-Slovak, and TikTok–-Spanish, indicating that keyword-level sampling variability plays a substantial role in uncertainty estimates.

Under joint uncertainty, intervals are widest overall, but the relative ranking of combinations remains largely unchanged. This suggests that the cross-platform and cross-language differences are robust to both annotation and retrieval uncertainty.

Overall, $P_{\text{total}}$ measures the share of mis/disinformation within the full content corpus, while $P_{\text{restricted}}$ gives the proportion only in relation to content for which a meaningful distinction between misleading and legitimate information can be made. Both measures show similar comparative patterns across platforms and languages. Therefore, deciding which indicator to use depends on individual preference or the specific use case. 

\clearpage
\section{Joint uncertainty results from pilot runs}   \label{app-D}

This appendix reports pilot results of the joint uncertainty procedure. As shown in Table~\ref{tab:by_language}, prevalence estimates are obtained with $B_{\text{kw}} = B_{\text{post}} = S \in {100, 200, 300}$ to assess the stability of the estimates as the number of draws increases. The results indicate that the joint uncertainty procedure produces stable estimates even at lower numbers of iterations. Increasing the number of draws does not alter the results significantly.

\begin{table}[htbp]
\centering
\caption{$P_{\text{restricted}}$ in (\%) by language -- Joint Uncertainty pilot runs.}
\label{tab:by_language}
\small
\begin{tabular}{@{}l rr rr rr@{}}
\toprule
& \multicolumn{2}{c}{100 Iterations}
& \multicolumn{2}{c}{200 Iterations}
& \multicolumn{2}{c}{300 Iterations} \\
\cmidrule(lr){2-3} \cmidrule(lr){4-5} \cmidrule(lr){6-7}
Language
& Est. & 95\% CI
& Est. & 95\% CI
& Est. & 95\% CI \\
\midrule

French   
& 19.8 & [16.7--23.1] 
& 19.6 & [16.0--22.6]
& 19.7 & [16.3--23.2] \\

Polish   
& 6.6 & [4.0--9.2]	 
& 6.5 & [4.0--8.9]
& 6.6 & [4.0--9.0] \\

Slovak   
& 8.9 & [5.5--16.3]
& 8.9 & [5.7--16.1]
& 8.9 & [5.7--16.1] \\

Spanish  
& 5.0 & [3.2--7.2]	 
& 4.9 & [3.1--6.7]
& 5.0 & [3.3--6.8] \\

\bottomrule
\end{tabular}
\end{table}

\clearpage
\section{$P_{\text{restricted}}$ estimates by platform language}   \label{app-E}

This appendix provides an illustration of $P_{\text{restricted}}$ estimates by platform-language reported in Section~\ref{plt-lang-resuls} of the manuscript.

\begin{figure}[h]
    \centering

   \includegraphics[width=0.98\textwidth, height=0.7\textheight]{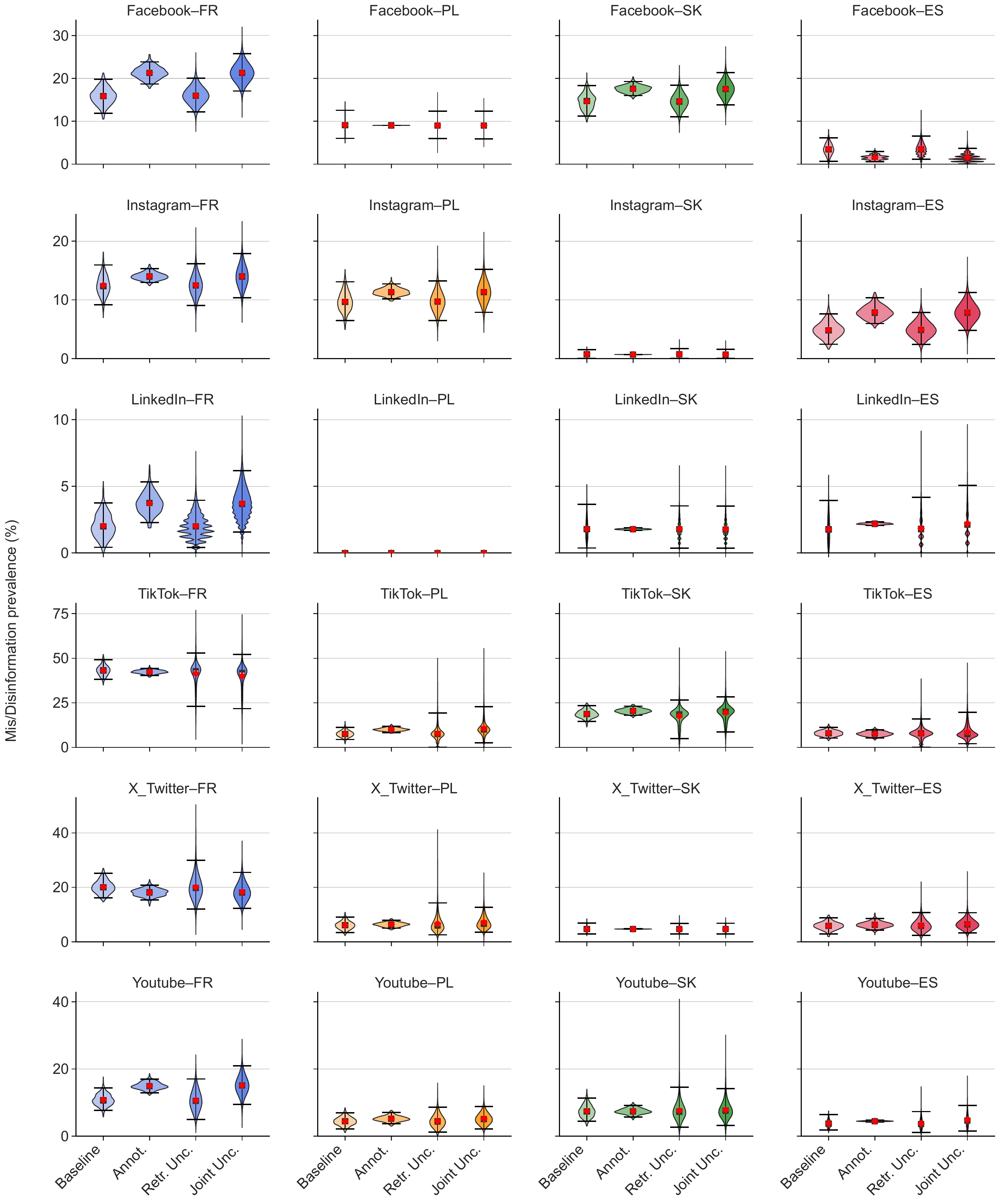}
    \caption{$P_{\text{restricted}}$ estimates and 95\% confidence intervals by platform and language under four uncertainty assumptions: baseline (Wilson), annotation uncertainty, retrieval uncertainty, and joint uncertainty. Internal markers provide a statistical summary: red diamonds for the mean, white dashes for the median, thick bars for the Interquartile Range, and capped whiskers for the 2.5th–97.5th percentiles. To ensure visual clarity and computational efficiency, the plots were generated using a downsampled dataset of 1,000 observations per method per platform-language combination.} 
    \label{fig:per_plat_lang}
\end{figure}

\section{LLM Prompt} \label{app-Prompt}

Full prompt used for large language model (LLM)-based content filtering is given below.

\begin{quote}\ttfamily
You are an expert fact-checker analyzing social media content for accuracy and information quality. Be objective, rigorous, thorough, and free from political or ideological bias.\\

INPUT\\
-- -- --\\
Content: [Content]\\
-- -- --\\
DEFINITIONS\\

\begin{itemize}
  \item [-] Important issues: topics likely to affect health/safety, civic participation/policy, personal or public economic decisions, or environmental risk, especially within the past 5 years; enduring medical guidance (e.g., vaccines, treatments) counts even if older.
  \item [-]Factual claim: a verifiable statement about the world (explicit or strongly implied).
\end{itemize}

DECISION PROCESS (apply in order)\\
\begin{itemize}
  \item [-]Extract the subject and any verifiable claims present in the visible text; if no claim is present, classify IRRELEVANT.
  \item [-]If claims exist, check whether they concern health/medicine/public health, public policy/government/elections/social issues, public safety/current events, economics/consumer decisions, or environment/climate.
  \item [-]Apply the impact test: would a reasonable reader change beliefs or actions in those domains after this post? Does the potential impact limited to just a small group of locality or it could affect the beliefs or actions of people from all over the world, with a stronger emphasis on the countries in Europe?
\end{itemize}

CLASSIFY AS\\

\begin{itemize}
    \item [-] RELEVANT if the content makes or implies verifiable claims that could shape reader beliefs or actions about important issues, including: 
    \subitem Health/medicine/public health (conditions, treatments, vaccines, diagnostics, supplements, risks, benefits, side effects).
    \subitem Public policy/government/elections/social issues.
    \subitem Public safety/current events/economics/finance decisions.
    \subitem Environmental risks/climate impacts.
    \subitem Satire/memes that assert real-world factual claims.
    \subitem Personal health stories that generalize, advise, compare treatments, cite efficacy/risks/side effects, or recommend actions.
    \subitem  Conspiracy/“alternative history” narratives that assert real events or institutional actions/cover-ups.
    \subitem Promotions/events only if they contain or repeat a verifiable claim or explicit prescriptive advice about an important issue (not merely “learn more” or logistics).
   \subitem  Personal health specificity: “Personal health stories are RELEVANT when they generalize or advise (efficacy, risk, side effects, comparisons, recommendations); donation or emotive appeals without such claims are IRRELEVANT”.
  \subitem  Institutional decisions: “Treat concrete policy changes, regulatory/enforcement actions, public safety alerts, economic measures, and institutional governance decisions (including sports bodies) as RELEVANT when claims are specific and verifiable”.
  \subitem Inflammatory or inciting content: Label as RELEVANT any post that harasses, dehumanizes, or calls for (or implicitly encourages) harassment, bullying, or violence against individuals, groups, organizations, or countries, even when framed as opinion, sarcasm, or lacks an explicit verifiable claim.
 \subitem Controversial/conspiracy insinuations: Label as RELEVANT any post that hints at, implies, or questions narratives about public‑impact domains or institutions (e.g., government, elections, public health, safety, economy, environment), including conspiracy theories or insinuations of cover‑ups, even when phrased as speculation, questions, memes, or slogans without concrete evidence in the post body.
 \subitem Historical events/stories which have strong influence on today's socio-political situations as well as people's political, religious beliefs.
\end{itemize}

\begin{itemize}
    \item [-]IRRELEVANT if it lacks such claims or impact, including:
    \subitem Historical/cultural/technology history without present-day health/safety/policy/economic implications.
  \subitem Pure entertainment, creative fiction, fandom.
  \subitem Personal diaries that are purely emotive or descriptive without advice, generalization, or broader factual assertions.
  \subitem Promotions/logistics/donation appeals with no factual claim or prescriptive advice (e.g., dates, tickets, “donate now”).
  \subitem Deleted/inaccessible content.
  \subitem Link teasers or extremely short posts that do not state a concrete verifiable claim in the visible text.
  \subitem Ceremonial/history carve-out: “Awards, tributes, obituaries, and historical/cultural posts are IRRELEVANT unless they assert current policy/safety/health/economic implications or generalizable claims that could change reader behavior”.
  \subitem Posts that clearly mention people, places, or events that are verifiably very local and that locality is entirely outside of Europe which we can say categorically have no impact on the lives of anyone outside that locality.
  \subitem Posts that are promotional in nature/intent (jobs, company policies, corporate events etc.). They could be touching on important issues and impacts but should be classified as IRRELEVANT since they are promotional in intent.

\end{itemize}

TIE-BREAKER IMPACT TEST\\

If uncertain, ask: “Would a reasonable reader change beliefs or actions related to health/safety/civic/economic/environmental decisions after this post?” If yes → RELEVANT; else → IRRELEVANT.\\

EXCEPTIONS:\\

\begin{itemize}
    \item [-] The two categories "incitement/harassment" and "controversy/conspiracy insinuations" are RELEVANT even if the visible text contains no concrete, verifiable claim, headline, or quantitative detail, because they can still shape beliefs and behaviors related to safety, civic trust, and public discourse.
    \item [-]Short posts or hashtags-only posts can be considered RELEVANT if the topics mentioned inferred from the visible words fall into the important categories/issues.
\end{itemize}

---\\

OUTPUT (JSON)\\

\{\ "classification": "RELEVANT" | "IRRELEVANT",
  "justification": "1–2 sentences naming the key claim(s) and the impact domain (e.g., public health, policy, safety, economics) and why it does or does not affect important issues"\}\ \\

EXAMPLES\\

\begin{itemize}
      \item [-] History of photography thread comparing daguerreotype vs. collodion with no current implications → IRRELEVANT (educational history without public-impact claims).
      \item [-] Personal cancer story: “My chemo regimen X works better than Y; avoid Y due to severe side effects” → RELEVANT (anecdotal but generalizable medical advice).
     \item [-]  Pure obituary/celebrity illness update with no advice or policy context → IRRELEVANT (life event without a public-impact claim).
     \item [-] Event promo: “Seminar shows vaccine Z causes infertility—register now” → RELEVANT (verifiable health claim plus prescriptive action).
    \item[-] Event promo: “Join us Friday—details in bio” → IRRELEVANT (logistics only; no claim).
    \item [-] Donation appeal: “Donate €10 to candidate Q” → IRRELEVANT (civic action request without factual claim).
    \item [-] Conspiracy/alt-history: “Empire T was erased from history by institution U” → RELEVANT (asserts real-world historical claims affecting civic trust).
    \item [-]Safety alert: “Scammers impersonate Bank W via SMS; don’t click links” → RELEVANT (public safety advice with a verifiable pattern).
    \item [-]Link teaser: “Trump recadre Poutine link” → IRRELEVANT (no visible claim).- History: “Les invasions mongoles… Source: Britannica” → IRRELEVANT (historical narrative without present implications).
    \item [-]Donation appeal: “Lina… cancer… donnez ici” → IRRELEVANT (no treatment claim or advice).
    \item [-]Safety alert: “Attention à cette nouvelle arnaque… vident votre compte” → RELEVANT (verifiable public safety claim and prescriptive warning).
     \item [-]Local policy: “Cagnes-sur-Mer: radars anti-bruit pour réduire les nuisances” → RELEVANT (municipal measure affecting environment/public policy).
    \item [-]Consumer health promo: “Déodorant compatible grossesse \& allaitement… 48h protection, sans alcool” → RELEVANT (safety/efficacy claims for a health-sensitive population).
     \item [-]Sports governance: “MLS interdit le garde du corps de Messi” → RELEVANT (institutional rule and safety rationale).
    \item [-]Economic policy/aid: “Leasing social à 100 € revient à partir de septembre” → RELEVANT (consumer/economic decision impact).
     \item [-]Local events/people: "The head of a Colombian town speaks about the rising crime rate in the first quarter of the year in his town." → IRRELEVANT eventhough it is an issue of importance, the impact is limited to the village in Colombia.
    \item [-] Historical figures/stories: "Hitler was a painter and during this period of youth, he learned his oratory skills." → RELEVANT . Stories like this could contain only facts and statements but because the topic is about a very influential figures which still get a significant following in the current socio-political climate, it is considered still relevant.
  \end{itemize}
\end{quote}

\end{appendices}


\bibliography{sn-bibliography}

@article{Fallis,
  title={What is disinformation?},
  author={Fallis, Don},
  journal={Library trends},
  volume={63},
  number={3},
  pages={401--426},
  year={2015},
  publisher={Johns Hopkins University Press}
}

@techreport{EDMO_second_report,
  title={Structural Indicators of the Code of Practice on Disinformation: The 2nd EDMO report},
  author={Nenadić, Iva and Brogi, Elda and Bleyer-Simon, Konrad and Reviglio, Urbano}, 
  year={2024},
  institution={European Digital Media Observatory}
}

@online{EC_CodePracticeDisinformation,
  author       = {{European Commission}},
  title        = {The 2022 Code of Practice on Disinformation},
  year         = {2025},
  url          = {https://test2.disinfocode.eu/wp-content/uploads/2023/09/code-of-practice-on-disinformation-september-22-2023.pdf},
  note         = {Accessed: 2025-12-12},
  organization = {European Commission},
  howpublished = {Online policy page}
}

@article{grabowskarole,
  title={The Role of the European Digital Media Observatory (EDMO) in Countering Disinformation in the European Union},
  author={Grabowska, Marta},
  journal={Poland’s Experience in Combating Disinformation},
  pages={71}
}

@misc{sfreport,
  title        = {Measuring the State of Online Disinformation in Europe on Very Large Online Platforms First report of the SIMODS project (Structural Indicators to Monitor Online Disinformation Scientifically)},
  author       = {Vincent, Emmanuel and Carniel, Bastien and Crisan, Daiana},
  year         = {2025},
  month        = sep,
  day          = {29},
  publisher    = {Science Feedback},
  url          = {https://science.feedback.org/wp-content/uploads/2025/09/SIMODS-Report-1.pdf},
  note         = {Accessed: 2025-12-15}
}

@article{ecker2024misinformation,
  title={Misinformation poses a bigger threat to democracy than you might think},
  author={Ecker, Ullrich and Roozenbeek, Jon and Van Der Linden, Sander and Tay, Li Qian and Cook, John and Oreskes, Naomi and Lewandowsky, Stephan},
  journal={Nature},
  volume={630},
  number={8015},
  pages={29--32},
  year={2024},
  publisher={Nature Publishing Group UK London}
}

@article{denniss2025social,
  title={Social media and the spread of misinformation: infectious and a threat to public health},
  author={Denniss, Emily and Lindberg, Rebecca},
  journal={Health promotion international},
  volume={40},
  number={2},
  pages={daaf023},
  year={2025},
  publisher={Oxford University Press US}
}

@article{glover2025fake,
  title={Fake News? Quantifying the prevalence of misinformation related to scoliosis on the Tik Tok Social Media Platform},
  author={Glover, Banahene and Datcu, Anne-Marie and Meyer, Macy and Lachmann, Emily and McIntosh, Amy and Johnson, Megan and Brooks, Jaysson T},
  journal={Journal of the Pediatric Orthopaedic Society of North America},
  pages={100207},
  year={2025},
  publisher={Elsevier}
}

@inproceedings{scarano2025election,
  title={Election Polls on Social Media: Prevalence, Biases, and Voter Fraud Beliefs},
  author={Scarano, Stephen and Vasudevan, Vijayalakshmi and Samory, Mattia and Yang, Kai-Cheng and Yang, JungHwan and Grabowicz, Przemyslaw A},
  booktitle={Proceedings of the International AAAI Conference on Web and Social Media},
  volume={19},
  pages={1771--1785},
  year={2025}
}

@article{ding2025evolvedetector,
  title={EvolveDetector: Towards an evolving fake news detector for emerging events with continual knowledge accumulation and transfer},
  author={Ding, Yasan and Guo, Bin and Liu, Yan and Jing, Yao and Yin, Maolong and Li, Nuo and Wang, Hao and Yu, Zhiwen},
  journal={Information Processing \& Management},
  volume={62},
  number={1},
  pages={103878},
  year={2025},
  publisher={Elsevier}
}

@article{quelle2025lost,
  title={Lost in translation: using global fact-checks to measure multilingual misinformation prevalence, spread, and evolution},
  author={Quelle, Dorian and Cheng, Calvin Yixiang and Bovet, Alexandre and Hale, Scott A},
  journal={EPJ Data Science},
  volume={14},
  number={1},
  pages={22},
  year={2025},
  publisher={Springer Berlin Heidelberg}
}

@inproceedings{nielsen2022mumin,
  title={Mumin: A large-scale multilingual multimodal fact-checked misinformation social network dataset},
  author={Nielsen, Dan S and McConville, Ryan},
  booktitle={Proceedings of the 45th international ACM SIGIR conference on research and development in information retrieval},
  pages={3141--3153},
  year={2022}
}

@article{liu2025systematic,
  title={A systematic review of machine learning approaches for detecting deceptive activities on social media: Methods, challenges, and biases},
  author={Liu, Yunchong and Shen, Xiaorui and Zhang, Yeyubei and Wang, Zhongyan and Tian, Yexin and Dai, Jianglai and Cao, Yuchen},
  journal={International Journal of Data Science and Analytics},
  pages={1--26},
  year={2025},
  publisher={Springer}
}

@online{codeConduct,
  author    = {{European Commission}},
  title     = {Code of Conduct on Disinformation},
  year      = {2025},
  month     = feb,
  url       = {https://digital-strategy.ec.europa.eu/en/library/code-conduct-disinformation},
  note      = {Shaping Europe’s digital future}
}

@article{brogi2024code,
  title={From the code of practice to the code of conduct? Navigating the future challenges of disinformation regulation},
  author={Brogi, Elda and De Gregorio, Giovanni},
  journal={Journal of Media Law},
  volume={16},
  number={1},
  pages={38--46},
  year={2024},
  publisher={Taylor \& Francis}
}

@inproceedings{awal2022muscat,
  title={MUSCAT: Multilingual rumor detection in social media conversations},
  author={Awal, Md Rabiul and Nguyen, Minh Dang and Lee, Roy Ka-Wei and Choo, Kenny Tsu Wei},
  booktitle={2022 IEEE International Conference on Big Data (Big Data)},
  pages={455--464},
  year={2022},
  organization={IEEE}
}

@inproceedings{pranesh2021looking,
  title={Looking for COVID-19 misinformation in multilingual social media texts},
  author={Pranesh, Raj Ratn and Farokhnejad, Mehrdad and Shekhar, Ambesh and Vargas-Solar, Genoveva},
  booktitle={European Conference on Advances in Databases and Information Systems},
  pages={72--81},
  year={2021},
  organization={Springer}
}

@article{khare2019relevancy,
  title={Relevancy identification across languages and crisis types},
  author={Khare, Prashant and Burel, Gregoire and Alani, Harith},
  journal={IEEE Intelligent Systems},
  volume={34},
  number={3},
  pages={19--28},
  year={2019},
  publisher={IEEE}
}

@article{quelle2023lost,
  title={Lost in translation--multilingual misinformation and its evolution},
  author={Quelle, Dorian and Cheng, Calvin and Bovet, Alexandre and Hale, Scott A},
  journal={arXiv preprint arXiv:2310.18089},
  year={2023}
}

@article{panchendrarajan2024claim,
  title={Claim detection for automated fact-checking: A survey on monolingual, multilingual and cross-lingual research},
  author={Panchendrarajan, Rrubaa and Zubiaga, Arkaitz},
  journal={Natural Language Processing Journal},
  volume={7},
  pages={100066},
  year={2024},
  publisher={Elsevier}
}

@article{hale2024analyzing,
  title={Analyzing misinformation claims during the 2022 Brazilian general election on WhatsApp, Twitter, and Kwai},
  author={Hale, Scott A and Belisario, Adriano and Mostafa, Ahmed Nasser and Camargo, Chico},
  journal={International Journal of Public Opinion Research},
  volume={36},
  number={3},
  pages={edae032},
  year={2024},
  publisher={Oxford University Press UK}
}

@article{liesenfeld2025legal,
  title={The legal significance of independent research based on article 40 DSA for the management of systemic risks in the digital services act},
  author={Liesenfeld, Anna},
  journal={European Journal of Risk Regulation},
  volume={16},
  number={1},
  pages={184--196},
  year={2025},
  publisher={Cambridge University Press}
}

@article{rogers2021marginalizing,
  title={Marginalizing the mainstream: How social media privilege political information},
  author={Rogers, Richard},
  journal={Frontiers in big Data},
  volume={4},
  pages={689036},
  year={2021},
  publisher={Frontiers Media SA}
}

@article{park2022beyond,
  title={Beyond performative transparency: lessons learned from the EU Code of Practice on Disinformation},
  author={Park, Kirsty and Culloty, Eileen},
  journal={AoIR Selected Papers of Internet Research},
  year={2022}
}

@article{fletcher2018measuring,
  title={Measuring the reach of" fake news" and online disinformation in Europe},
  author={Fletcher, Richard and Cornia, Alessio and Graves, Lucas and Nielsen, Rasmus Kleis},
  journal={Australasian Policing},
  volume={10},
  number={2},
  pages={25--33},
  year={2018},
  publisher={Australasian Institute of Policing Melbourne, Vic.}
}

@article{ozcelik2025detecting,
  title={Detecting misinformation on social media using community insights and contrastive learning},
  author={Ozcelik, Oguzhan and Toraman, Cagri and Can, Fazli},
  journal={ACM Transactions on Intelligent Systems and Technology},
  volume={16},
  number={2},
  pages={1--27},
  year={2025},
  publisher={ACM New York, NY}
}

@inproceedings{rashid2025probabilistic,
  title={A Probabilistic Reasoning Framework to Detect Fake News on Social Media},
  author={Rashid, Mehreen and Anber, Fabliha and Samiullah, Md and Ahmed, Chowdhury Farhan and Leung, Carson K and Pazdor, Adam GM},
  booktitle={2025 19th International Conference on Ubiquitous Information Management and Communication (IMCOM)},
  pages={1--8},
  year={2025},
  organization={IEEE}
}

@article{Wilson01061927,
author = {Edwin B. Wilson},
title = {Probable Inference, the Law of Succession, and Statistical Inference},
journal = {Journal of the American Statistical Association},
volume = {22},
number = {158},
pages = {209--212},
year = {1927},
publisher = {Taylor \& Francis},
doi = {10.1080/01621459.1927.10502953}
}

@article{agresti1998approximate,
  title={Approximate is better than “exact” for interval estimation of binomial proportions},
  author={Agresti, Alan and Coull, Brent A},
  journal={The American Statistician},
  volume={52},
  number={2},
  pages={119--126},
  year={1998},
  publisher={Taylor \& Francis}
}

@article{brown2001interval,
  title={Interval estimation for a binomial proportion},
  author={Brown, Lawrence D and Cai, T Tony and DasGupta, Anirban},
  journal={Statistical science},
  volume={16},
  number={2},
  pages={101--133},
  year={2001},
  publisher={Institute of Mathematical Statistics}
}

@article{suarez2021prevalence,
  title={Prevalence of health misinformation on social media: systematic review},
  author={Suarez-Lledo, Victor and Alvarez-Galvez, Javier},
  journal={Journal of medical Internet research},
  volume={23},
  number={1},
  pages={e17187},
  year={2021},
  publisher={JMIR Publications Toronto, Canada}
}

@article{kbaier2024prevalence,
  title={Prevalence of health misinformation on social media—challenges and mitigation before, during, and beyond the COVID-19 pandemic: scoping literature review},
  author={Kbaier, Dhouha and Kane, Annemarie and McJury, Mark and Kenny, Ian},
  journal={Journal of Medical Internet Research},
  volume={26},
  pages={e38786},
  year={2024},
  publisher={JMIR Publications Toronto, Canada}
}

@report{trustlab2023pilot,
  author = {{TrustLab}},
  year = {2023},
  title = {Code of Practice on Disinformation -- A comparative analysis of the prevalence and sources of disinformation across major social media platforms in Poland, Slovakia and Spain},
  institution = {University of Cambridge},
  type         = {Report},
  url          = {https://www.trustlab.com/resources/codeofpractice-disinformation#report},
  note         = {Accessed: 2026-02-18}
}

@inproceedings{artstein2009bias,
  title={Bias decreases in proportion to the number of annotators},
  author={Artstein, Ron and Poesio, Massimo},
  booktitle={Proceedings of FG-MoL 2005: The 10th conference on Formal Grammar and The 9th Meeting on},
  volume={139},
  year={2009}
}

\end{document}